\documentclass[manuscript,nonacm]{acmart}

\AtBeginDocument{%
  }

\setcopyright{acmlicensed}
\copyrightyear{2018}
\acmYear{2018}
\acmDOI{XXXXXXX.XXXXXXX}

\acmConference[Conference acronym 'XX]
{Make sure to enter the correct conference title from your rights confirmation email}
{June 03--05, 2018}
{Woodstock, NY}

\acmISBN{978-1-4503-XXXX-X/2018/06}

\usepackage{graphicx}
\usepackage{wrapfig}
\usepackage{subcaption}
\usepackage{xcolor}
\usepackage{tikz}
\usetikzlibrary{shapes.geometric, arrows}
\usepackage{comment}
\usepackage{todonotes}
\usepackage{float}

\tikzstyle{startstop} = [
    rectangle,
    rounded corners,
    minimum width=3cm,
    minimum height=1cm,
    text centered,
    draw=black
]

\tikzstyle{io} = [
    trapezium,
    trapezium stretches=true,
    trapezium left angle=70,
    trapezium right angle=110,
    minimum height=0.7cm,
    text centered,
    draw=black,
    inner sep=3pt
]

\tikzstyle{process} = [
    rectangle,
    text centered,
    minimum height=0.7cm,
    draw=black,
    inner sep=3pt
]

\tikzstyle{decision} = [
    diamond,
    text centered,
    draw=black
]

\tikzstyle{arrow} = [thick, ->, >=stealth]

\definecolor{matrixblue}{RGB}{31,119,180}

\newcommand{\matrixcell}[3]{%
    \pgfmathsetmacro{\intensity}{min(#3*7.5, 100)}
    \ifnum#3=0
        \node[
            draw=gray!70,
            line width=1.1pt,
            fill=white,
            minimum size=11mm,
            inner sep=0pt
        ] at (#1,#2) {{\LARGE 0}};
    \else
        \node[
            draw=gray!70,
            line width=1.1pt,
            fill=matrixblue!\intensity!white,
            minimum size=11mm,
            inner sep=0pt
        ] at (#1,#2) {{\LARGE #3}};
    \fi
}

\begin{document}


\title{An Affordable Vibrotactile Surface for Tactile Figure Recognition: Exploration Strategies of Blind and Sighted Users}


\author{Uday Nath Gunda}
\email{p20240026@goa.bits-pilani.ac.in}
\affiliation{%
  \institution{BITS Pilani, K. K. Birla Goa Campus}
  \city{Goa}
  \country{India}
}

\author{Arun Raman}
\email{araman@goa.bits-pilani.ac.in}
\affiliation{%
  \institution{BITS Pilani, K. K. Birla Goa Campus}
  \city{Goa}
  \country{India}
}

\author{Parthasarathy PD}
\email{p20210042@goa.bits-pilani.ac.in}
\affiliation{%
  \institution{BITS Pilani, K. K. Birla Goa Campus}
  \city{Goa}
  \country{India}
}


\begin{abstract}
Refreshable tactile displays can make graphics accessible to blind
and low-vision people, but high-resolution actuator arrays remain
expensive. We study whether a low-cost vibrotactile surface can
convey simple figures, and how they are read. Nine participants
(four blind; five sighted, tested blindfolded then open-eyed)
identified seven figures by touch, with video and timing analysis
of their exploration. Blind and blindfolded-sighted readers
achieved comparable accuracy, but blind participants were about
three times faster, using a continuous, rhythmic ``gliding''
strategy---reading through motion---whereas sighted participants
searched in halting, pause-heavy passes; among the sighted,
more pausing accompanied errors. Recognition failures were
concentrated in specific figures (a diagonal was misread even
with vision), locating the bottleneck in representation rather
than resolution. We contribute an empirical characterization of
tactile reading strategies and design implications for low-cost
refreshable tactile displays and affordable tactile representation.
\end{abstract}


\maketitle


\section{Introduction}

Modern educational practice increasingly relies on visual representations such as interactive plots, diagrams, and animations~\cite{menendez2026learning}. These representations are largely inaccessible to blind and low-vision (BLV) learners, creating an urgent need for technologies and methods that include them in the educational ecosystem~\cite{doore2023multimodality}. This need is compounded by how such material is delivered in practice: BLV learners are frequently supported by sighted teachers whose role includes producing and mediating tactile representations of visual content which is a labour-intensive process constrained by the time, tools, and resources available to them \cite{phutane2022tactile}. Any practical tactile technology must therefore also serve the sighted educators who operate it.

Traditional tactile learning relies on static representations such as embossed diagrams. These are non-interactive and cannot be easily updated or reprinted, making them unsuitable for interactive learning \cite{vidal2007graphical}. Refreshable tactile displays can render text, shapes, and dynamic content, but remain expensive, bulky, and mechanically complex, limiting their reach in {resource-constrained settings} \cite{bhatnagar2023analysis}. Screen readers convert text to audio effectively but they provide limited access to spatial, temporal, and can't provide data visualizations \cite{thompson2023chart, zong2022rich}. Against this backdrop, vibrotactile stimulation offers a promising path \cite{choi2012vibrotactile}. It leverages the skin's mechanoreceptors to deliver information. By trading precise point-actuation for low-cost, off-the-shelf vibration motors, a vibrotactile surface can potentially approximate refreshable tactile output at a fraction of the cost. However, the class of figures and spatial relationships that a vibrotactile surface can and cannot convey reliably remains an open question. 

This paper reports the first iteration (along with a preceding zero-th iteration) of an ongoing program on a low-cost vibrotactile surface for BLV learners. In both iterations, the surface renders discrete figures and arrangements of active pixels\footnote{In this paper, we place a vibration motor in a housing or a casing, we refer to it as a pixel.} read by touch, and we ask two linked questions: which figures an affordable, low-resolution surface can reliably convey, and how blind and sighted people read them. The zero-th iteration was with a 5x3 surface where its resolution confined figures that we studied to compact glyphs such as digits and Braille alphabets. Its evaluation exposed limitations that reshaped the design. The present iteration is a reworked, 10x10 version which addresses some limitations of the first, while the added resolution expands the representable class to open geometric forms with extended strokes and diagonals.

Designing such surfaces is non-trivial because vibration is not a point phenomenon and energy propagates through the substrate or a material. Consequently, a user may perceive unintended vibration from interference between neighbouring pixels, distorting the represented information. There are other issues such as actuator size and signal damping upon touch. The zero-th iteration, mentioned in the previous paragraph, made another difficulty concrete: that the actuators could not be individually replaced. The first iteration, on which this paper is mostly based, reworks the pixel and its housing to address these challenges (Section \ref{sec:design}). One key innovation is that of a load-decoupling cap that keeps the motor responsive under finger pressure. These changes, alongside the increased resolution of 10 x 10 let us broaden the investigation on which figures the surface can reliably render, and how blind and sighted people read them.


The current iteration costs approximately \$100. Each tactile pixel of the 10×10 grid is made of a cylindrical vibration motor embedded in a 3D-printed casing. The surface renders figures composed of straight-line strokes; its resolution precludes smooth curves, bounding the representable class. In a study with four blind and five sighted participants, we examine both what the surface conveys and which figures are recognized, and which are confused and how participants read it, through their manual exploration of the surface. Sighted participants read each figure first without vision, then with vision available to guide exploration. This two-stage protocol lets us separate limits of the display from limits of exploration, and sighted readers also reflect the educators who may mediate tactile material in practice. Blind participant read the figures in a single stage.


We found that the surface renders simple axis-aligned line figures reliably but not the plus or the diagonal. We discover that the blind and sighted participants follow distinct exploration strategies likely influenced by their experience with Braille and computer keyboard respectively. There is no significant difference in the recognition accuracy between blind and blindfolded sighted participants but Blind participants are about three times faster. This paper makes the following contributions:
\begin{enumerate}
\item The iterative design of an affordable vibrotactile surface, including a novel load-decoupling pixel that keeps cylindrical vibration motors responsive under finger pressure, and a physical characterization of its rendering behaviour including inter-pixel isolation.
\item An empirical study with blind and sighted users characterizing which figures the surface reliably renders and the structure of figure confusions, distinguishing display limitations from exploration and perception.
\item An analysis of manual exploration strategies, derived from per-trial video, relating strategy to recognition success, including a comparison between blind and blindfolded-sighted readers.
\item Design implications for low-cost refreshable tactile displays and directions toward affordable tactile representation.
\end{enumerate}

\section{Related Work}
 
\subsection{Refreshable and Low-Cost Tactile Displays}

Traditional tactile graphics, such as embossed drawings and swell-paper, convey figures and text effectively, but they are static and often difficult to produce spatial information. Studies of tactile materials in educational settings have recognized both their value and the practical constraints in their creation availability, and reuse \cite{phutane2022tactile}. Refreshable tactile displays(RTDs) address some of these issues by generating the tactile content that can be mechanically refreshed to represent text, graphics, diagrams, and flowcharts. Commercially available RTDs like Graphiti and Dot Pad use arrays of individually controlled tactile pixels~\cite{orbitgraphiti, dotpad}. Some recent work has investigated the use of RTDs for accessible data visualizations and interactive data exploration suggesting that their refreshability can extend tactile exploration beyond fixed graphics ~\cite{holloway2024refreshable, reinders2025when, marriott2025vision}.
Despite these advantages, producing large arrays of controlling individual tactile pixels remains hard because of mechanical actuators. Magnepins uses electromagnetic actuation and off-the-shelf components to build a large refreshable tactile arrays at lower cost \cite{smiley2025magnepins}, while electrovibration provides another approach, producing tactile feedback through skin resistance without requiring mechanical moving pins \cite{bau2010teslatouch}. These developments reflect the trade-off between tactile resolution and scalability, which is a major factor in determining the cost of the system. Our work explores this trade-off from a different direction: using vibration motors as discrete tactile pixels. This reduces mechanical complexity of the display but introduces new perceptual and interaction challenges in interpreting low-resolution vibrotactile patterns.

\subsection{Vibrotactile Spatial Displays and Pattern Perception}

Vibrotactile spatial displays have a long history in assistive technology, beginning with the Optacon, which converted visual information captured by a camera into spatial patterns of vibration on the fingertip~\cite{linvill1966direct,bliss1970optical}. Subsequent studies showed that users can learn to recognize vibrotactile patterns and letters, with experienced users achieving relatively high reading performance~\cite{craig1977vibrotactile,craig1981tactile}. Reviews of vibrotactile display research have since highlighted the role of stimulus and display characteristics in shaping tactile perception and performance~\cite{choi2012vibrotactile}. 
{Studies of vibrotactile pattern recognition have also examined how pattern complexity and the spatial structure of the display affect discrimination. Horner et al. found that more complex vibrotactile patterns were more difficult to distinguish~\cite{horner1991effects}, while Garcia-Hernandez et al. examined how characteristics of the tactile array influence the perception of spatial patterns~\cite{garcia2010feeling}. Together, these studies point to a relationship between the physical resolution of a vibrotactile display, the complexity of the represented figure, and recognition performance. Most of this earlier work, however, focuses on relatively dense fingertip-scale arrays or interaction through a localized area of the skin. Our work examines a different setting: a large, low-resolution grid of independent vibration motors that can be explored freely with the hand. This allows us to study how recognizable spatial figures are when tactile information is distributed across a larger surface with substantially fewer active elements.}
 
\subsection{Tactile Pattern Recognition and Haptic Exploration} 

Studies on tactile spatial resolution have shown that the fingertip has only limited ability for distinguishing closely spaced features; therefore, features that are too close together may not be perceived as separate features by the user \cite{johnson1981tactile, vanboven1994limit}. At the same time, recent work with low-resolution tactile displays shows that meaningful graphical shapes can still be recognised when they are represented using discrete tactile elements \cite{bhatnagar2023pixelated}. These findings suggest that tactile recognition is not determined by spatial resolution alone, but also by how users interpret and connect the individual elements that form a pattern.
Exploration is therefore an important factor in tactile shape/pattern recognition. Lederman and Klatzky identified particular patterns of hand movements called exploratory procedures that allow participants to obtain various types of information through touch \cite{klatzky1992stages, lederman1987hand}. Subsequent work then demonstrated that haptic exploration can be in multiple stages, with users first gaining an overall understanding of an object before focusing on specific features \cite{klatzky1992stages}. Recent work on planar tactile shapes has similarly found that people use different hand movements, including contour following and scanning, when identifying shapes \cite{sadia2022exploration}. Together, this work demonstrates that both the spatial structure of a tactile pattern and the way it is explored contribute to recognition, a setting close to ours. We build on this work by coding participants' manual exploration of our surface from per-trial video and relating it to recognition success

\subsection{Tactile Ability in Blind and Sighted Users}

Whether blindness confers superior tactile ability is a longstanding and unresolved question that bears directly on our study design. Several studies report finer tactile acuity in blind readers than in sighted controls: \citet{vanboven2000tactile} found lower grating-orientation thresholds in blind Braille readers, and \citet{goldreich2003tactile} argued that this enhancement follows from visual deprivation itself, largely independent of Braille experience. Others attribute the advantage to accumulated tactile experience rather than deprivation, observing that blind and sighted participants can perform comparably once the sighted are given practice~\cite{grant2000tactile}. This debate motivates our design. Because blindfolded sighted people are not equivalent to blind people---their spatial cognition remains visually scaffolded, and they lack a lifetime of tactile reliance---we treat blindfolded sighted reading as a control condition rather than a proxy for blindness, and we include both blind and sighted participants so that their recognition and exploration can be compared directly.

\section{Zero-th Iteration and Redesign Rationale}
\label{sec:first-iteration}

We initiated our development with a 5x3 array and studied compact glyphs in the form of numeric digits and Braille as those can fit a small, low-resolution grid. We describe the zero-th iteration briefly to motivate the present design.

\subsection{The 5\texttimes3 Surface}

\begin{figure}[h]
    \centering
    \begin{subfigure}{0.33\textwidth}
        \centering
        \includegraphics[width=\linewidth]{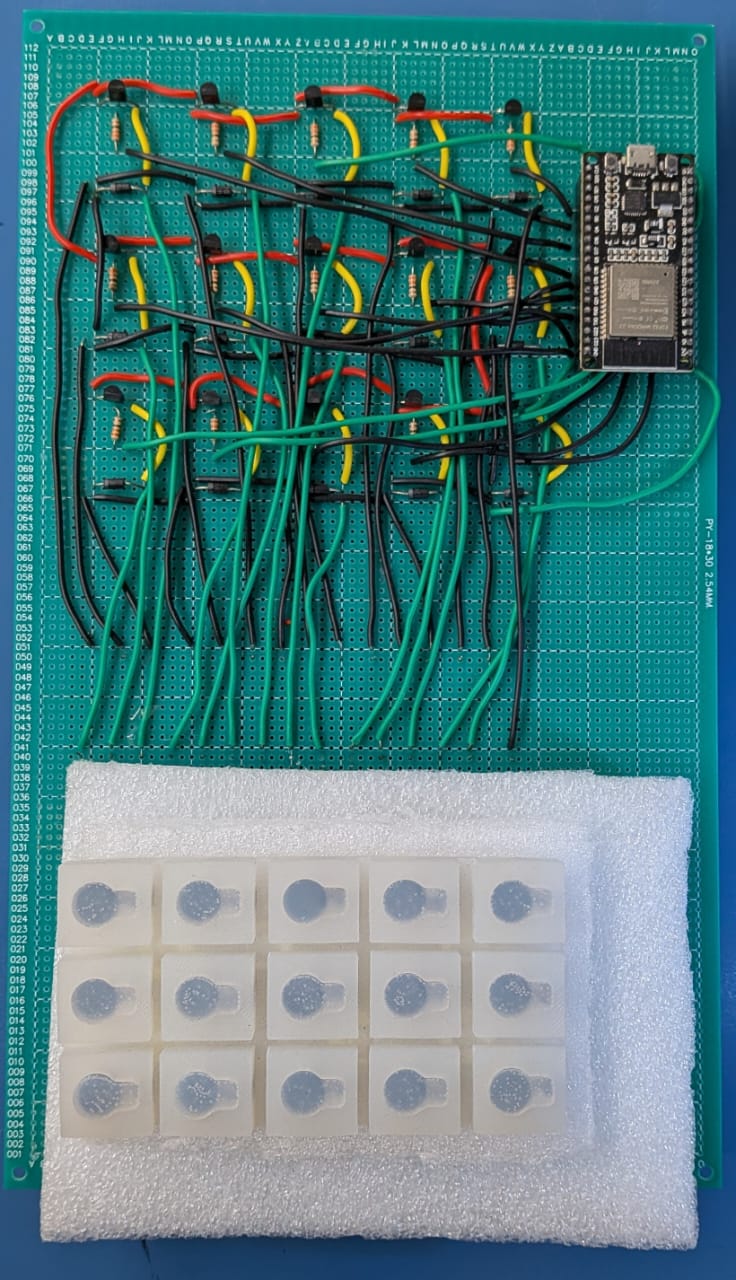}
        \caption{5 x 3 surface.}
        \Description{The image shows a single perfboard with two parts. The top part has a control circuit made with an ESP32 and other small components such as transistors, resistors, and diodes. This part controls the system. The bottom part has 15 small vibration motors. These motors are connected to the control circuit with wires and are used to create vibrations.}
    \end{subfigure}
    \hfill
    \begin{subfigure}{0.55\textwidth}
        \centering
        \includegraphics[width=\linewidth]{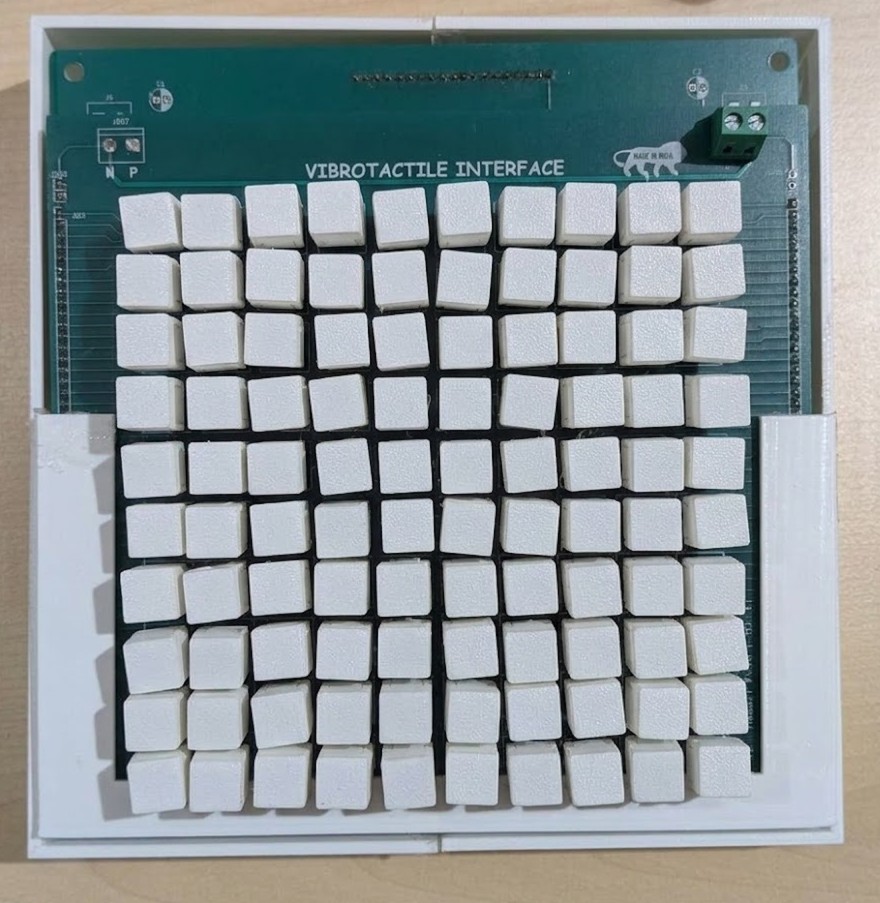}
        \caption{10 x 10 surface.}
        \Description{The image of the 10 × 10 vibrotactile surface consisting of 100 individually arranged 3D-printed pixels in a square grid. Each pixel contains a vibration motor and is separated from the neighbouring pixels by small gaps. The pixels are mounted within a PCB.}
    \end{subfigure}
    \caption{The evolution of the surface design from the zero-th to the first iteration. }
    \Description{}
\end{figure}

The zero-th iteration was a $5\times3$ array of eccentric rotating mass (ERM)
coin motors, each embedded in a cast poly-di-methyl-siloxane (PDMS) capsule and
driven by a microcontroller and transistor circuit. Digits and Braille
characters were rendered by activating the corresponding subset of motors. We
conducted a user study in which 9 sighted participants and one blind participant
identified the rendered glyphs by touch. The study established that recognition
was \emph{feasible}: participants were able to identify glyphs presented as
vibrotactile patterns. 

\subsection{Limitations and Design Responses}
The zero-th iteration exposed three limitations, which the present design
addresses. In several cases a single design change resolves more than one
limitation; we group them accordingly.

\paragraph{Resolution and cost.}
The coin motors imposed a coarse pixel pitch, confining the surface to small,
dense glyphs, and the cast PDMS capsules were expensive. The present iteration
adopts smaller-diameter cylindrical motors, which allow a finer pitch and hence
more pixels in the same footprint, and replaces PDMS with a low-cost,
3D-printable poly-lactic-acid (PLA) casing.
\begin{wrapfigure}[9]{l}{0.21\textwidth}
  \centering
    \includegraphics[width=0.95\linewidth]{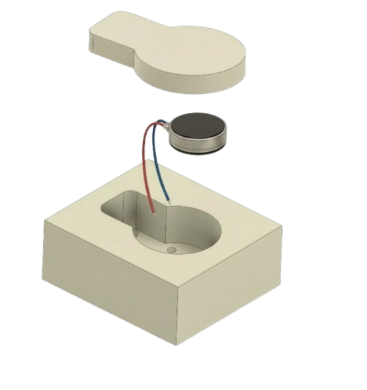}
    \Description{This figure shows the parts of one PDMS pixel separated to explain how they fit together. There is a PDMS base with a hollow space inside it. A small round vibration motor with two wires is placed inside this hollow space. A PDMS top cover is placed above the motor to close the pixel and keep the motor fixed in position.}
    \caption{Component assembly view for the 5 x 3 design.}
    \label{fig:assembly}
\end{wrapfigure}
\paragraph{Contact damping and motor stalling.}
In the 5x3 design, resting the hand on the surface damped the vibration,
so that patterns could not be discerned when the palm was in contact. The move
to cylindrical motors made this contact-sensitivity acute as a cylindrical ERM stalls under even light pressure on its vibrating end. We address this with a load-decoupling cap that receives vibration from the base of the motor while shielding the vibrating element from direct finger pressure (Section~\ref{sec:design}).

\paragraph{Inter-pixel isolation and pixel replacement.}
The first iteration's motors shared a single cast substrate, which both coupled
vibration between neighbouring pixels and made an individual motor impossible to
replace without de- and re-soldering fine wire---a repair inaccessible to BLV
users, as we noted in that iteration. The present design instead builds each
pixel as an independent unit that connects to the PCB through a berg-strip
(header) connector. Separating the pixels removes the shared cast substrate that
coupled neighbouring pixels, and the connector makes a pixel removable by hand.
Together with the removable cap, this is intended to let a BLV user replace a
damaged motor without tools or soldering---remove the cap, withdraw the motor,
and seat a new one. We designed for this affordance; evaluating whether BLV
users can perform the repair unaided is future work.

%
%
\begin{figure}[h]
    \centering
    \begin{subfigure}{0.45\textwidth}
        \centering
        \includegraphics[width=\linewidth]{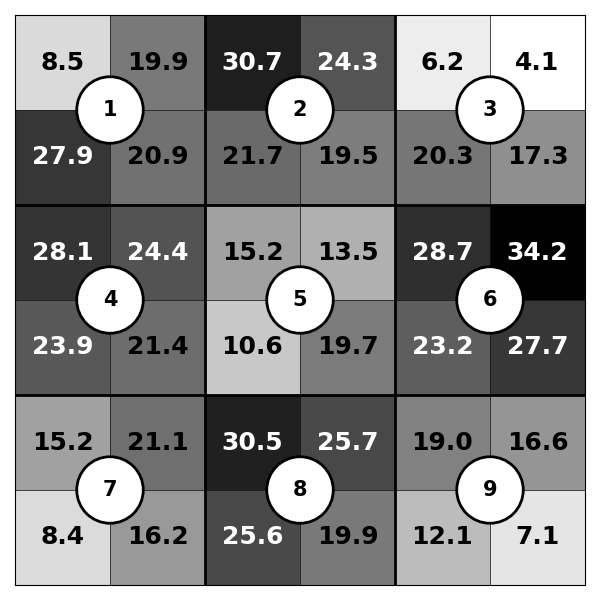}
        \caption{PSF with $5\times3$ Surface with motors 2, 4, 6 and 8 ON. Each pixel divided into subpixels for resolution.}
        \label{fig:psf}
        \Description{A 3×3 grid divided into nine labeled regions (1–9), each containing four numeric values representing measurements. Regions 2, 4, 6, and 8 correspond to active vibrating motors and show higher values compared to the other regions, indicating multiple vibration sources distributed across the grid, while the remaining regions show comparatively lower readings.}
    \end{subfigure}
    \hfill
    \begin{subfigure}{0.45\textwidth}
        \centering
        \includegraphics[width=0.9\linewidth]{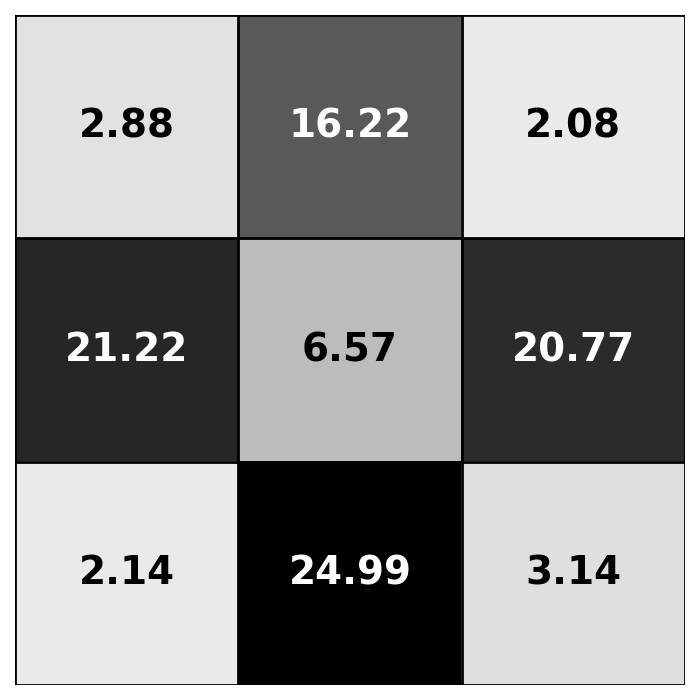}
        \caption{PSF with $10 \times 10$ surface}
        \Description{The figure shows the point spread function (PSF) measured on the 10 × 10 vibrotactile surface. The 3 × 3 grid represents the active pixel and its eight surrounding regions, with the measured values displayed in each region. The values are 2.88, 16.22, 2.08, 21.22, 6.57, 20.77, 2.14, 24.99, and 3.14, arranged from top-left to bottom-right. The central region has a value of 6.57, while the surrounding regions show values ranging from 2.08 to 24.99.}
        \label{psf_1}
    \end{subfigure}
    \caption{The point spread functions (PSF) for the designs.}
    \label{PSF}
    \Description{There are two images showing point spread function for both the iterations}
\end{figure}
\section{$10 \times 10$ Vibrotactile Interface}
\label{sec:design}

Guided by the limitations of the 5 x 3 prototype, we built a $10\times10$ interface that raises spatial resolution and improves repairability. A larger array is also
needed to render figures beyond compact glyphs--- geometric forms here, and
richer tactile data in future iterations ---which requires more, and more densely
packed, pixels.

\subsection{Mechanical Design}
\label{sec:mech}
\begin{wrapfigure}[10]{l}{0.23\textwidth}
  \centering
    \includegraphics[width=0.5\linewidth]{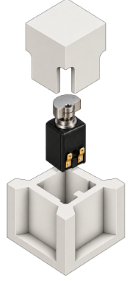}
    \caption{Component Assembly.}
    \Description{This figure shows the parts of one tactile pixel separated to explain how they fit together. There are two 3D-printed casing parts, one above and one below the motor. A small micro-vibration motor is placed between the two casing parts. The lower casing has a hollow space that holds the motor, while the upper casing is placed over the motor to close the pixel and keep the motor fixed in position. The black rectangular part in the middle is the micro-vibration motor, with a small metal rotating part on top and two electrical contacts at the bottom.}
    \label{fig:assembly}
\end{wrapfigure}
Each pixel had to satisfy three goals: a compact casing housing a small
cylindrical vibration motor, a reliable electrical connection without permanent
per-motor wiring, and replacement of an individual motor without soldering. We
used the cheaper wireless motor variant, which exposes metal terminals, and over
several fabrication iterations converged on a two-part casing: a \emph{bottom} part that holds the motor body and a \emph{top} cap that encloses it and forms the touch surface, joined by a sliding enclosure that keeps the assembly rigid while allowing tool-free removal of the motor. The final casing measures $11\times11\times16\,\text{mm}$ at a $2\,\text{mm}$ pitch, is 3D-printed in PLA, and is arranged in a $10\times10$ grid. Assembling each
pixel independently before integration removes almost all manual wiring of the
assembled surface.

\paragraph{Load decoupling.}
A cylindrical ERM motor stalls under even light pressure on its vibrating end,
which would damp the motor precisely when the surface is touched. The two-part
casing resolves this: the finger contacts the rigid top cap, held to the bottom
part by the sliding enclosure, so contact force is carried into the casing
rather than onto the motor. A $3\,\text{mm}$ clearance between the inner top of
the cap and the vibrating end of the motor ensures the finger never loads the
motor directly, while the cap stays coupled to the motor at its base and
therefore continues to transmit vibration to the fingertip. The same removable
cap makes a motor field-replaceable---withdraw the cap, remove the motor, seat a
new one, with no tools or soldering. We designed this to be performed by BLV
users themselves; evaluating whether they can do so unaided is future work.

\begin{figure}[h]
    \centering
    \begin{subfigure}{0.34\textwidth}
        \centering
        \includegraphics[width=\linewidth]{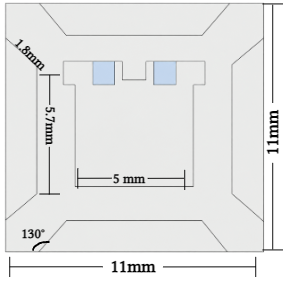}
        \caption{Top view of the base of the casing}
        \Description{This figure shows the top view of the bottom casing with its main dimensions. The base is a square in a shape with a width and length of 11 mm. A hollow space is provided in the center to hold the micro-vibration motor. The central space is about 5.7 mm deep and 5 mm wide at the bottom. The casing has angled sections at the corners, with an angle of 130°. The wall thickness around the central hollow space is about 1.8 mm. Two small rectangular areas are shown near the top of the central space for the motor connections.}
    \end{subfigure}
    \hfill
    \begin{subfigure}{0.30\textwidth}
        \centering
        \includegraphics[width=0.85\linewidth]{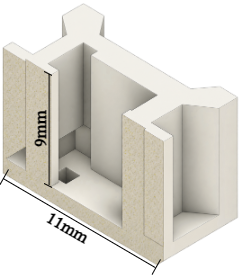}
        \caption{Cut section of the base}
        \Description{This figure shows a cut section of the bottom casing. The base is a square-shaped 3D-printed part with a width of 11 mm. The inside has a hollow space where the micro-vibration motor is placed. The inner space is 9 mm deep, with walls around it that support and protect the motor. A small opening is also provided at the bottom of the hollow space. The cut section shows the inside structure and the walls of the base clearly.}
    \end{subfigure}
    \hfill
    \begin{subfigure}{0.30\textwidth}
        \centering
        \includegraphics[width=0.85\linewidth]{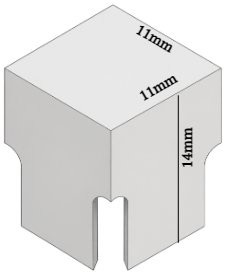}
        \caption{Top part}
        \Description{This figure shows the 3D view of the top part of the casing. The top part is a square-shaped 3D-printed casing with a length and width of 11 mm and a height of 14 mm. A narrow opening is provided at the bottom center of the casing. The sides have curved cutouts near the lower part. The top part is designed to fit over the micro-vibration motor and the bottom casing.}
    \end{subfigure}

    \caption{Specifications of an individual ``pixel''.}
    \Description{}
\end{figure}

\subsection{Hardware}
\label{sec:hw}

The interface separates the surface from its control electronics across two
stacked PCBs: a \emph{motor PCB} carrying the 100 cased motors in their spatial
$10\times10$ arrangement, and a \emph{control PCB} carrying the ESP32, 13 chained
74HC595 shift registers, and a per-motor MOSFET switch stage. The boards are
joined by four 25-pin Berg-strip connectors carrying the 100 motor-control
signals with a mechanical support for holding the top-PCB. Each motor mates with a female connector on the motor PCB, so a motor
can be added or removed without soldering to its terminals. Mounting each pixel
as an independent unit on a rigid PCB, rather than in a single cast block as in
the 5 x 3 iteration, removes the shared substrate that coupled neighbouring
pixels reduces inter-pixel vibration coupling relative to
the moulded design. The point spread functions (PSFs) of the two designs are shown in Fig. \ref{PSF}. The procedure used to calculate the PSF is in the Appendix. 

\subsection{Software}
\label{sec:sw}

The interface is controlled by the ESP32, programmed via the Arduino IDE. The
nine figures used in the study are predefined in firmware, each stored as the
list of motor numbers to activate. The ESP32 hosts
a local Wi-Fi web server through which the experimenter selects a figure; the
corresponding motor list is clocked out through the 74HC595 chain to switch the
appropriate MOSFETs, producing the spatial vibrotactile figure. 

\section{Research Questions}

We use the designed surface to render, as spatially distributed vibrotactile patterns, discrete rectilinear figures composed of horizontal, vertical, and diagonal strokes that a user has to explore and recognize by touch. Curved and circular forms are not in the scope as they cannot be approximated smoothly due to the low spatial resolution of the current design. We organize our investigation around four research questions:
 
\begin{description}
  \item[RQ1 (Rendering fidelity).] Which figures does the surface render as
  tactilely distinct and reliably recognizable, and which pairs of figures are confused at this resolution?
 
  \item[RQ2 (Exploration strategy).] What manual strategies do participants use
  to read a figure on the surface, and whether and how do these strategies
  relate to recognition success and reading time? 
 
  \item[RQ3 (Effect of visual access).] For sighted participants, how do
  recognition accuracy, reading time, and exploration change between the
  blindfolded and open-eyed stages, and which confusions persist even when
  vision is available?
 
  \item[RQ4 (Blind vs.\ sighted touch readers).] How do blind participants
  compare with blindfolded sighted participants---both reading purely by
  touch---in recognition accuracy, confusion structure, and exploration
  strategy? 
\end{description}
 
\noindent
RQ1 characterizes the display; RQ2 characterizes the reading process; RQ3 and
RQ4 isolate, respectively, the effect of vision and the effect of tactile
expertise on that process.

\section{User Study}
\subsection{Participants and Consent}
We recruited nine participants: four Braille-literate congenital blind (age between 30 and 40, 1 female and 3 males) and five sighted (age range 22 to 32, 3 females and 2 males) through purposive and convenience sampling. The four blind participants were recruited purposively as members of the target user population. Two out of the the five sighted participants were also recruited purposively as they had provided feedback for the previous design also (Section \ref{sec:first-iteration}). Table \ref{tab:participants} has more details. Sighted participants provided informed consent through a Google Form; blind participants gave explicit audio-recorded consent. Prior IRB approval was obtained from the researchers' institute for the study.

\begin{figure}[h]
    \centering
    \begin{subfigure}{0.65\textwidth}
        \centering
        \includegraphics[width=\linewidth]{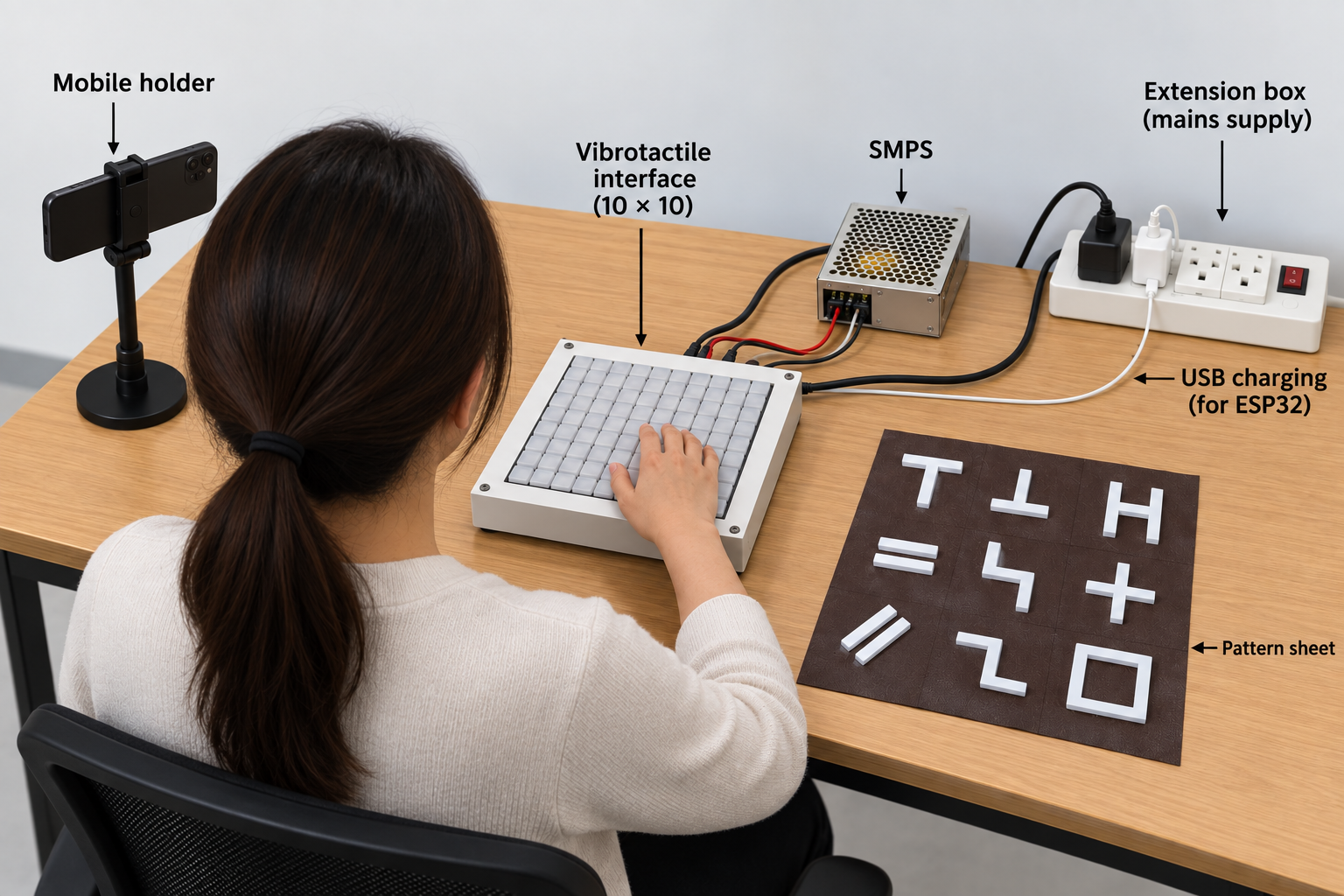}
        \caption{AI-generated illustration of the experimental setup used in our study. It also shows the 9 figures used in the study, as part of the reference sheet.}
       \Description{This figure shows the experimental setup used in the study. A person is seated at a table and interacting with a vibrotactile interface labelled 10 × 10. The interface is connected to an SMPS and an extension box that provides the main power supply. A USB cable is used for powering the ESP32. A mobile phone is mounted on a holder on the left side of the setup. A pattern sheet with different shapes is placed next to the interface for the user to follow during the experiment.}
       \label{fig:mobileplacement}
    \end{subfigure}
    \hfill
    \begin{subfigure}{0.33\textwidth}
        \centering
        \includegraphics[width=\linewidth]{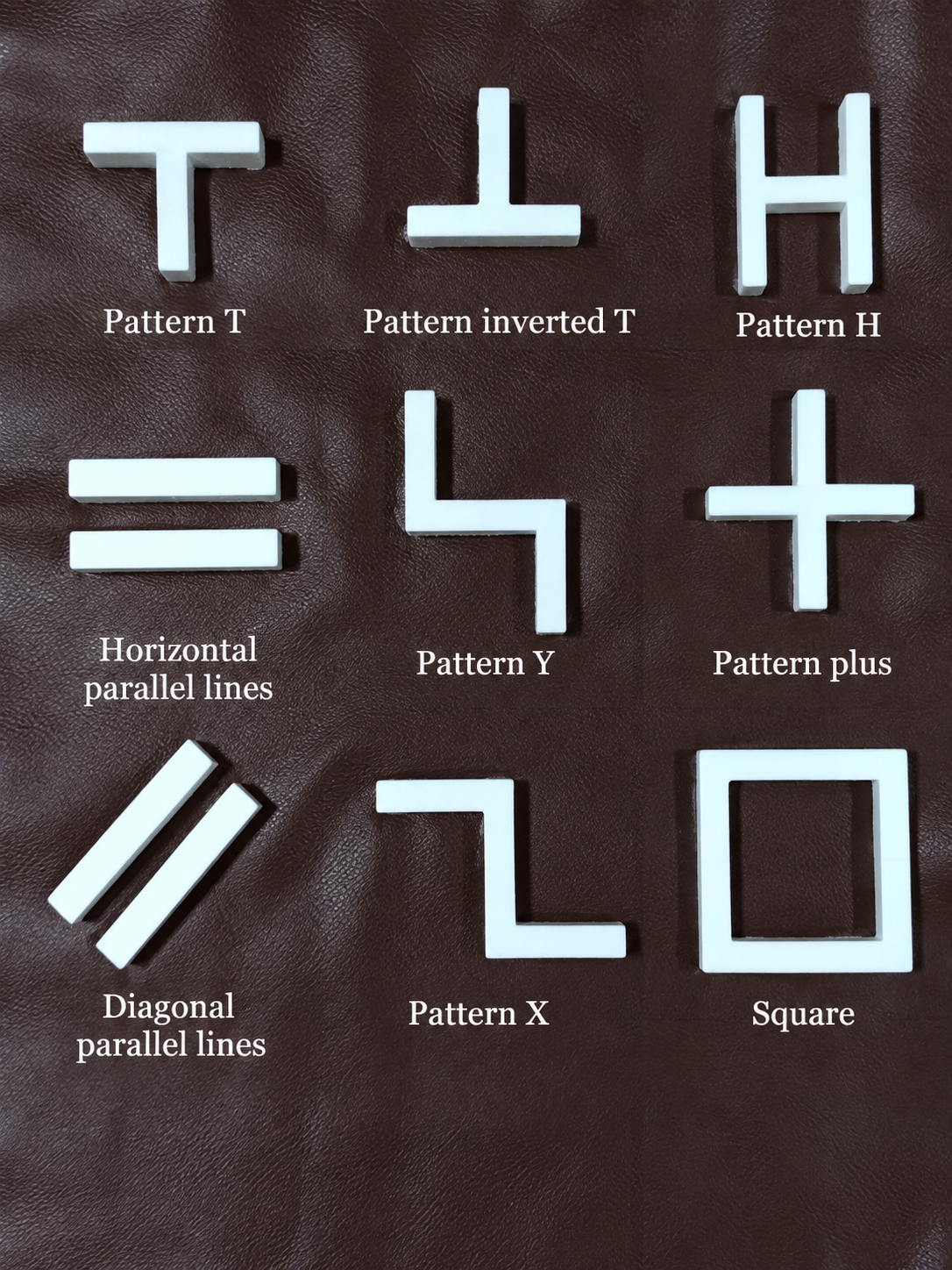}
        \caption{3D-printed patterns attached on a leather sheet.}
        \Description{A set of nine tactile patterns arranged in a 3 × 3 layout on a dark surface. The patterns are labeled Pattern T, Pattern inverted T, Pattern H, Horizontal parallel lines, Pattern Y, Pattern plus, Diagonal parallel lines, Pattern X, and Square. Each pattern is formed using 3D-Printed white elements, with the line-based patterns representing different orientations and geometric arrangements for tactile exploration.}
        \label{full}
    \end{subfigure}
    \caption{}
\end{figure}

\begin{table*}
  \caption{Participant Demographics and Previous Experience}
  \label{tab:participants}
  \begin{tabular}{cclllll}
    \toprule
    Participant & Age & Gender & BLV & Braille Literacy & Previous Experience with Haptic Devices \\
    \midrule
    P1 & 22 & Male & --- & No & No \\
    P2 & 40 & Female & Congenital & Yes & Yes \\
    P3 & 40 & Male & Congenital & Yes & Yes \\
    P4 & 29 & Female & --- & --- & Yes, involved in 5$\times$3 user study \\
    P5 & 30 & Male & --- & --- & Yes, involved in 5$\times$3 user study \\
    P6 & 22 & Female & --- & No & No \\
    P7 & 30 & Male & Congenital & Yes & Yes \\
    P8 & 31 & Male & Congenital & Yes & Yes \\
    P9 & 21 & Female & --- & No & No \\
    \bottomrule
  \end{tabular}
\end{table*}


\subsection{Study Methodology.} 
The study used nine rectilinear figures rendered on the surface — two for training (T and inverted T) and seven for testing. Each figure was available as a 3D printed shape pasted on a sheet (Fig. \ref{fig:mobileplacement}), which participants could read by touch or by sight. During training, the researcher sometimes guided the participant's hand to the matching figure on the reference sheet to establish the correspondence between the rendered vibration pattern and its figure. 

Following training, sighted participants completed the study in two stages: first blindfolded, then open-eyed. Blind participants completed a single session. The seven test figures were arranged into four fixed permutations. At the start of testing, each participant selected a number from 1 to 4, which determined the order in which figures were presented to them. For sighted participants, the same selected sequence was repeated in both stages.
On each trial, a single figure was rendered on the surface; the participant explored it by touch and responded by identifying the corresponding figure on the reference sheet. A trial continued until the participant responded or chose to give up. The researcher disclosed whether a response was correct only if the participant asked. After every third trial, participants were asked whether the vibration was causing any discomfort and were offered rest. Sighted participants then repeated the identical sequence in the open-eyed stage. Vibrations of some motors were visible with naked eye. For each trial we recorded the presented figure, the participant's response, its correctness, and the total trial time. All trials were video-recorded with the participant's hands in view (see Fig. \ref{fig:mobileplacement} for camera positioning). {In total, we have 116 recorded videos (testing + training).} 

\section{Qualitative Analysis}
\subsection{Methodology}
\begin{figure}[h]
    \centering
    \begin{subfigure}{0.45\textwidth}
        \centering
        \includegraphics[width=\linewidth]{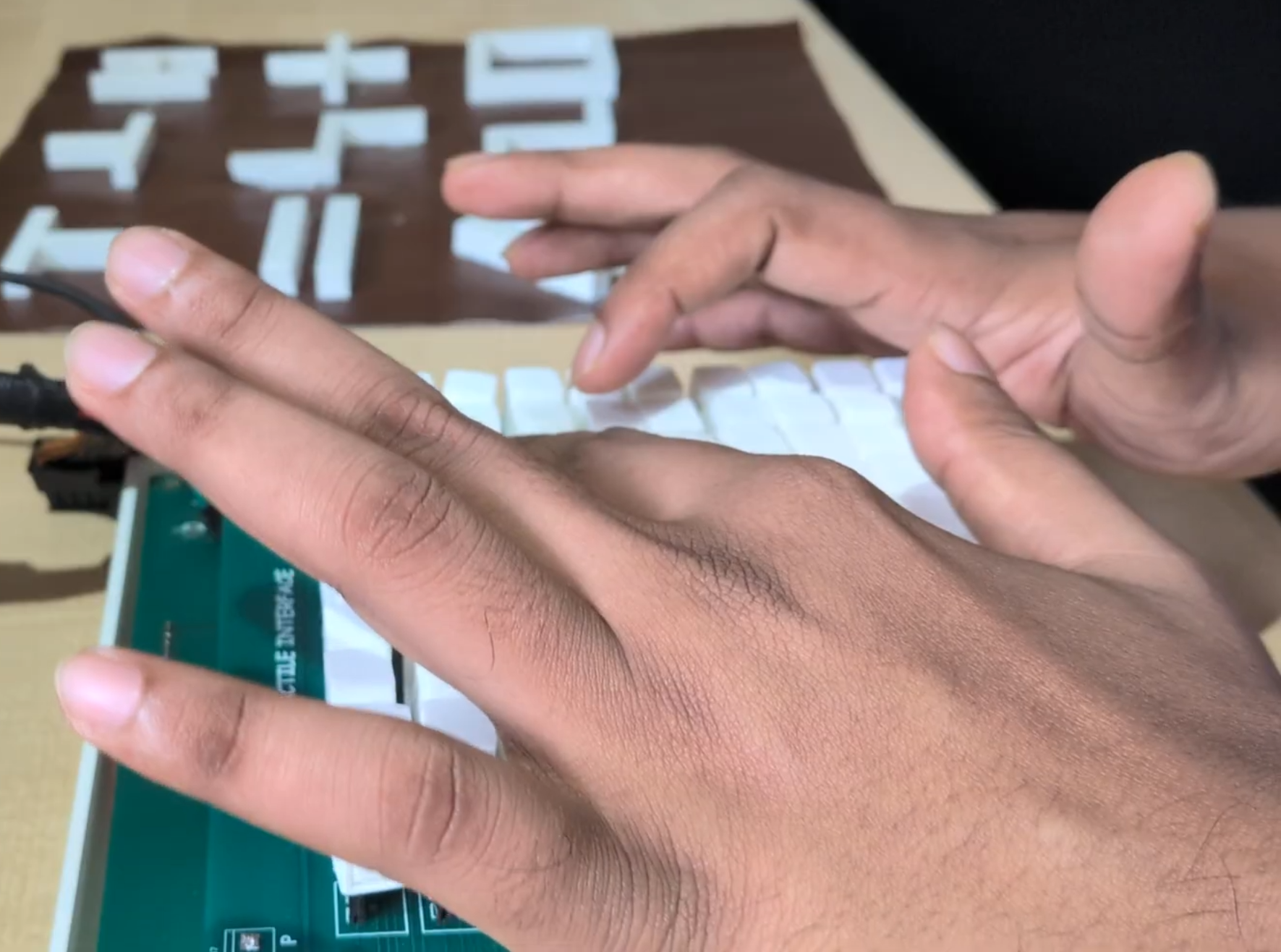}
        \caption{User obstructing the camera while interacting with device}
        \Description{This figure shows a user interacting with the tactile pattern display while their hand blocks the camera view. One hand is placed over the tactile display, while the other hand is also close to the device.This shows how the user's hand can obstruct the camera while interacting with the device.}
        \label{fig:obstruction}
    \end{subfigure}
    \hfill
    \begin{subfigure}{0.45\textwidth}
        \centering
        \includegraphics[width=\linewidth]{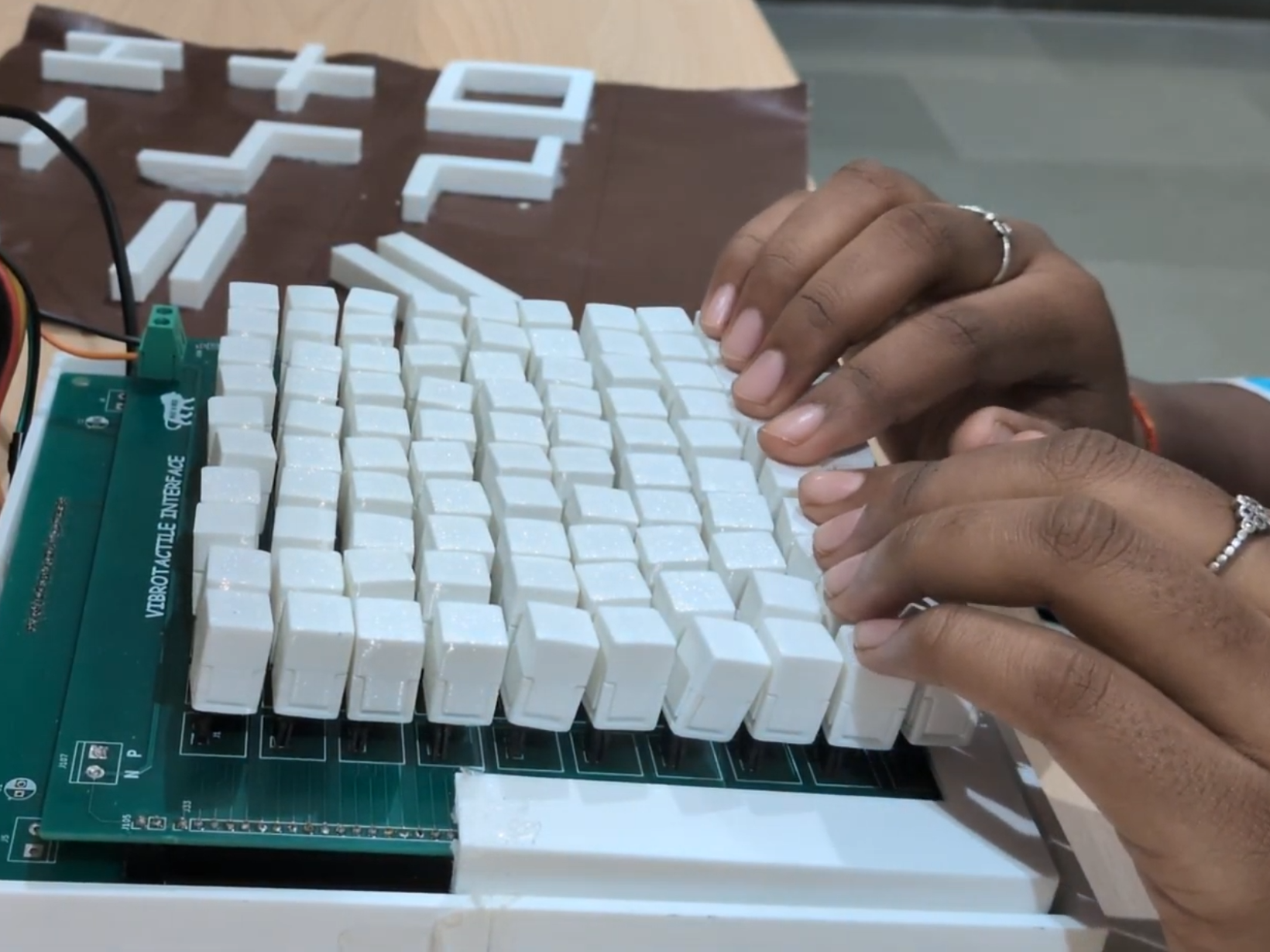}
        \caption{Participant using 8 fingertips as in a keyboard.}
        \Description{This figure shows a user using eight fingers to interact with a tactile display like a keyboard. The user places both hands over the raised blocks and uses multiple fingers to press and feel different parts of the display.}
        \label{full}
    \end{subfigure}
    \caption{}
\end{figure}

We conducted structured video analysis to characterize participants’ temporal and manual exploration behaviour. Two coders independently coded each testing trial video, with each trial corresponding to one pattern presentation. Temporal measures included time spent interacting with the reference sheet, vibrotactile surface, and both simultaneously (we use this for the analysis in Section \ref{sec:results}). Manual exploration was categorically characterized by the number of fingers used by each hand (0–5), hand use (left, right, or both), scanning strategy (discrete jumps or continuous swipes) and contact type: tip, full finger or palm for exploration.  The coding captured only the presence and absence of these features. Following independent coding of nine videos, the coders met and refined their interpretation of the coding scheme. We excluded the training trials from the analysis resulting in 98 trials for analysis. After both coders completed the independent coding, we assessed inter-rater reliability using Cohen's $\kappa$ for the coded variables (Table \ref{CohenK}). There is substantial agreement between the raters on all features except in the contact type feature where is moderate but at the border of substantial. Agreement is confounded by camera positioning: in a number of trials the user's hand partially occluded the view of the surface (Fig. \ref{fig:obstruction} is an illustration), obscuring fine movements and forcing coders to infer them (they did not resort to a default guess). 




\begin{figure}[h]
    \centering
    \begin{subfigure}{0.30\textwidth}
        \centering
        \includegraphics[width=\linewidth]{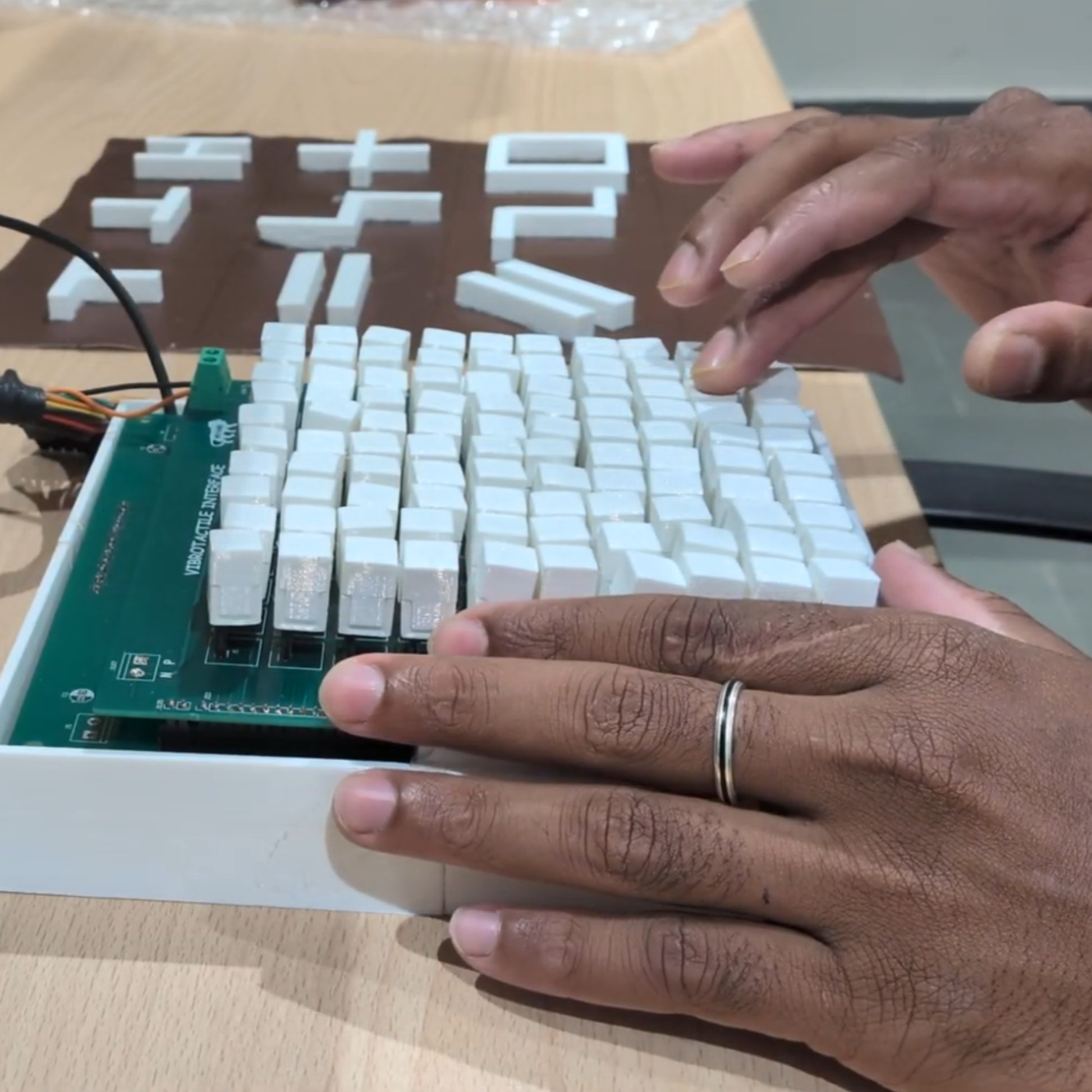}
        \caption{Finger Tip}
        \Description{This figure shows a person using their fingertip to interact with the vibrotactile interface.The person touches the raised blocks with their fingers to feel the tactile pattern.}
    \end{subfigure}
    \hfill
    \begin{subfigure}{0.30\textwidth}
        \centering
        \includegraphics[width=\linewidth]{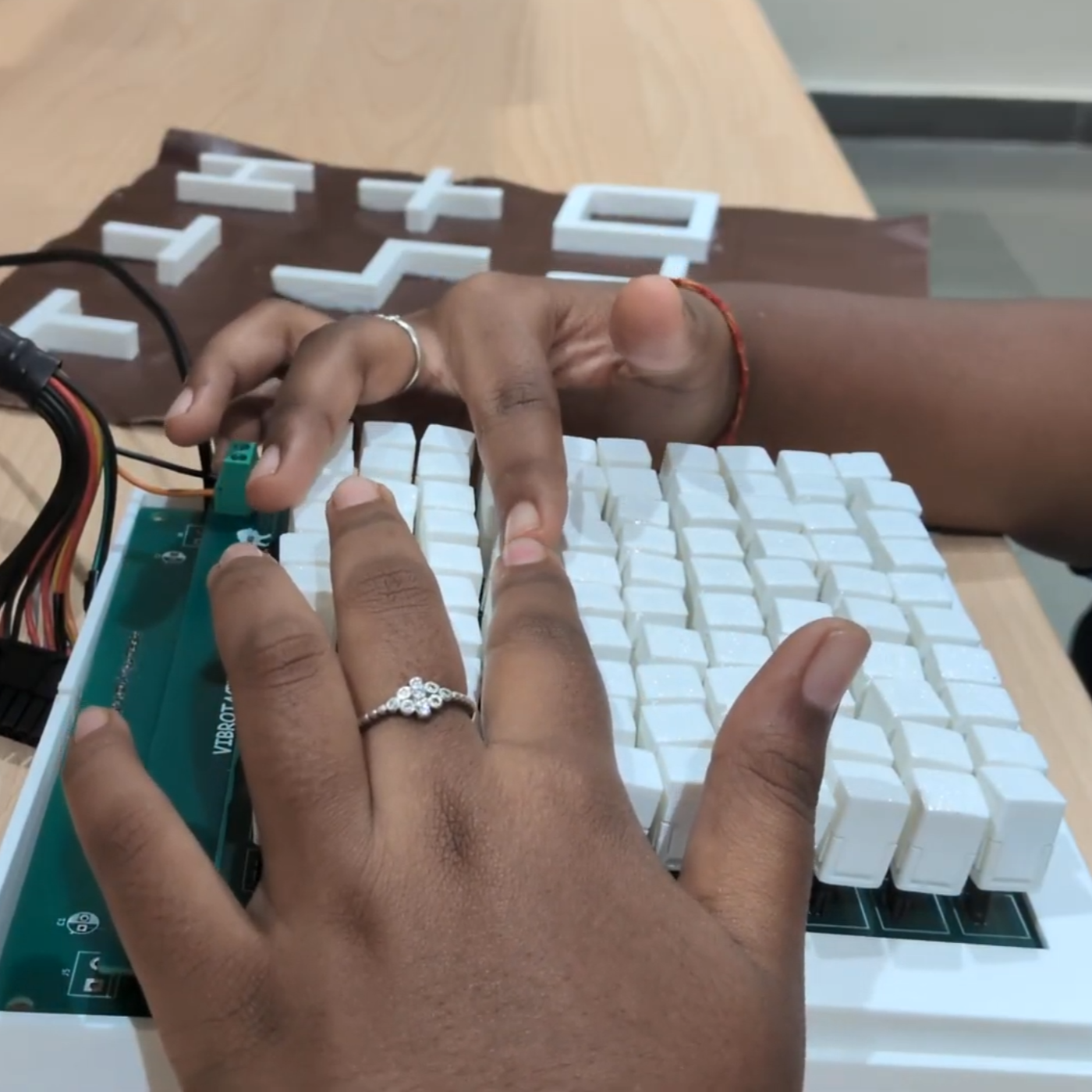}
        \caption{Full Finger}
        \Description{This figure shows a person using their full fingers to interact with a tactile pattern display. The person places their whole finger that covers multiple tactile pixels.}
    \end{subfigure}
    \hfill
    \begin{subfigure}{0.35\textwidth}
        \centering
        \includegraphics[width=\linewidth]{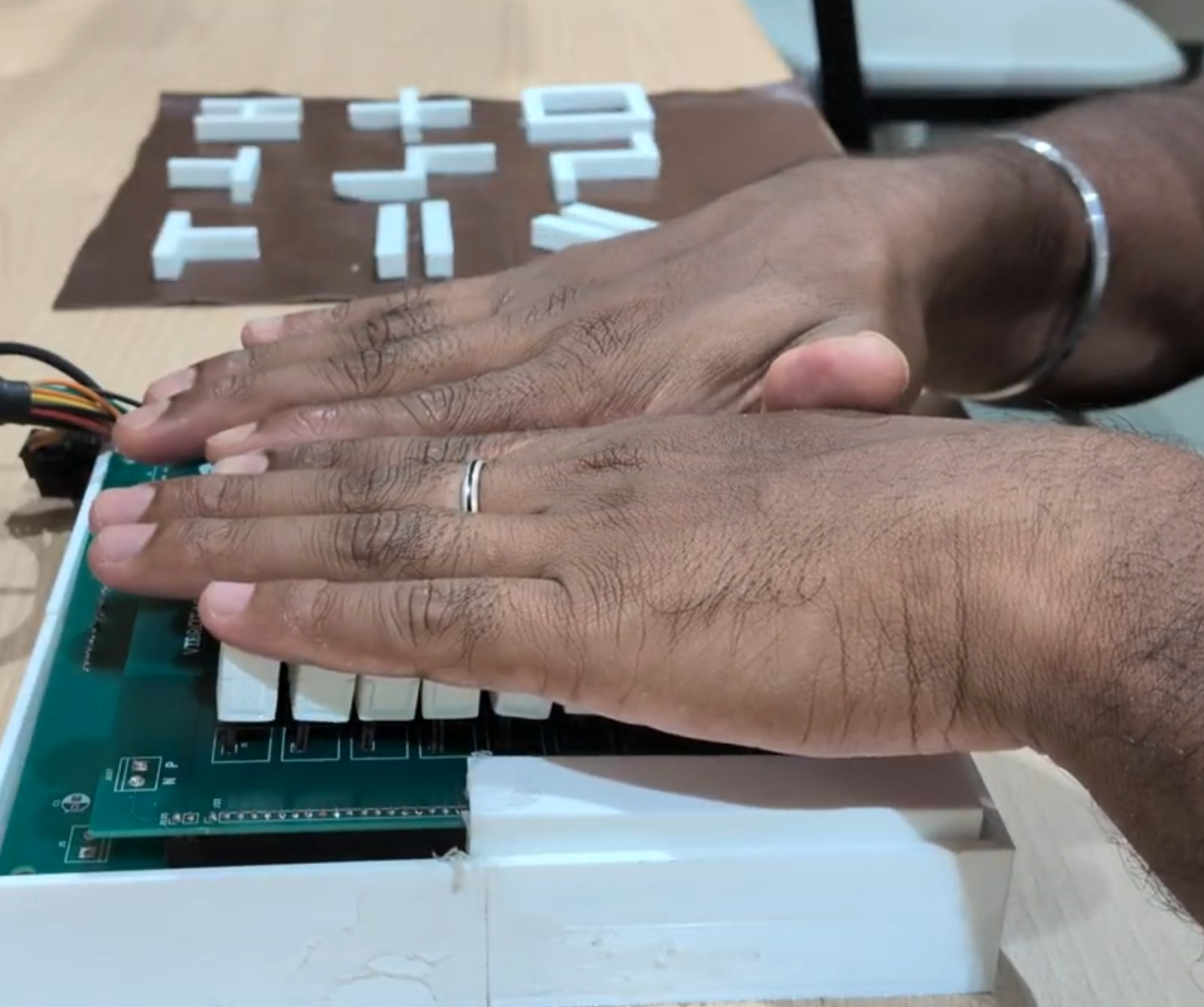}
        \caption{Palm}
        \Description{This figure shows a person using their palm to interact with a tactile pattern display. The person places their palm over the raised blocks to feel the tactile pattern.}
    \end{subfigure}
    \label{fig: HandInteraction}
    \caption{Users Interaction.}
    \Description{}
\end{figure}

\begin{table*}[t]
    \centering

    \begin{minipage}[t]{0.48\textwidth}
        \centering
        \caption{Classification accuracy for each pattern.}
        \label{tab:pattern_accuracy}
        \begin{tabular}{lc}
            \toprule
            Pattern & Accuracy (\%) \\
            \midrule
            Horizontal Parallel Lines & 92.86 \\
            Pattern Y                 & 86.67 \\
            Pattern X                 & 73.33 \\
            H                         & 64.29 \\
            Square                    & 64.29 \\
            Plus                      & 64.29 \\
            Diagonal Parallel Lines   & 42.86 \\
            \bottomrule
        \end{tabular}
        \Description{his table shows the classification accuracy for seven tactile patterns. The accuracy values are: Horizontal Parallel Lines, 92.86\%; Pattern Y, 86.67\%; Pattern X, 73.33\%; H, 64.29\%; Square, 64.29\%; Plus, 64.29\%; and Diagonal Parallel Lines, 42.86\%.}
    \end{minipage}
    \hfill
    \begin{minipage}[t]{0.48\textwidth}
        \centering
        \caption{Cohen's $\kappa$ values for different features.}
        \label{tab:cohen_kappa}
        \begin{tabular}{lc}
            \toprule
            Feature & Cohen's $\kappa$ \\
            \midrule
            \#Fingers - Left    & 0.673 \\
            \#Fingers - Right   & 0.628 \\
            Contact Type (T, F, P)      & 0.601 \\
            Hands (L, R, B)            & 0.795 \\
            Strategy (D or C)           & 0.716 \\
            \bottomrule
        \end{tabular}
        \label{CohenK}
    \end{minipage}
    \Description{This table shows Cohen’s kappa values for seven different features. The features and their values are: Fingers–Left, 0.673; Fingers–Right, 0.628; Type, 0.601; Hands, 0.795; Scan, 1.000; Coverage, 1.000; and Contact, 0.716.}
\end{table*}

\begin{table}[t]
  \caption{Hand-contact modes used to read the surface, by condition (number of
  trials). Each trial was coded for the presence of fingertip (T), full-finger
  (F), and/or palm (P) contact; rows are the observed combinations. Results from Coder 1.}
  \label{tab:contactmode}
  \small
  \begin{tabular}{lcccc}
    \toprule
    Contact & Blind & Sighted & Sighted & Total \\
     mode   &       & Blindfold & Open-Eye &      \\
    \midrule
    T          &  1 &  0 &  1 &  2 \\
    F          &  2 &  0 &  0 &  2 \\
    P          &  0 &  0 &  0 &  0 \\
    T,\,F      & 11 & 15 & 25 & 51 \\
    T,\,P      &  0 &  0 &  1 &  1 \\
    F,\,P      &  6 &  4 &  0 & 10 \\
    T,\,F,\,P  &  8 & 16 &  8 & 32 \\
    \midrule
    Total      & 28 & 35 & 35 & 98 \\
    \bottomrule
  \end{tabular}
\end{table}

\subsection{Analysis and Results}
Table~\ref{tab:contactmode} summarizes the categorization of the contact mode by the primary coder. Reading was almost exclusively finger-based: fingertip contact appeared in 86 of 98 trials and full-finger contact in 95, whereas the palm was never used on its own. The dominant mode was combined fingertip-and-full-finger exploration (T, F; 51 trials), with the palm added in a further 32 (T, F, P). Rarer patterns were idiosyncratic to individuals rather than group tendencies: the two full-finger-only trials came from one blind participant, and six of the ten fingertip-absent (F, P) trials from another. The clearest condition effect was in palm use, which was common under blindfolded touch (20/35 sighted-blindfolded; 14/28 blind) but roughly halved once vision was available (9/35), as sighted participants shifted toward the simpler fingertip-plus-full-finger mode (T, F rising from 15 to 25 trials); fingertip use was near-universal open-eyed (35/35).

We also recorded the number of fingers contacting the surface per hand but that should be read as breadth of contact rather than sustained exploration (participants often engaged extra fingers only briefly). The two hands were near-symmetric overall---median four fingers each, with 62 of 98 trials identical across hands. This parameter was most affected by the occlusion of the camera.



\begin{figure}[h]
        \centering
        \includegraphics[width=0.7\linewidth]{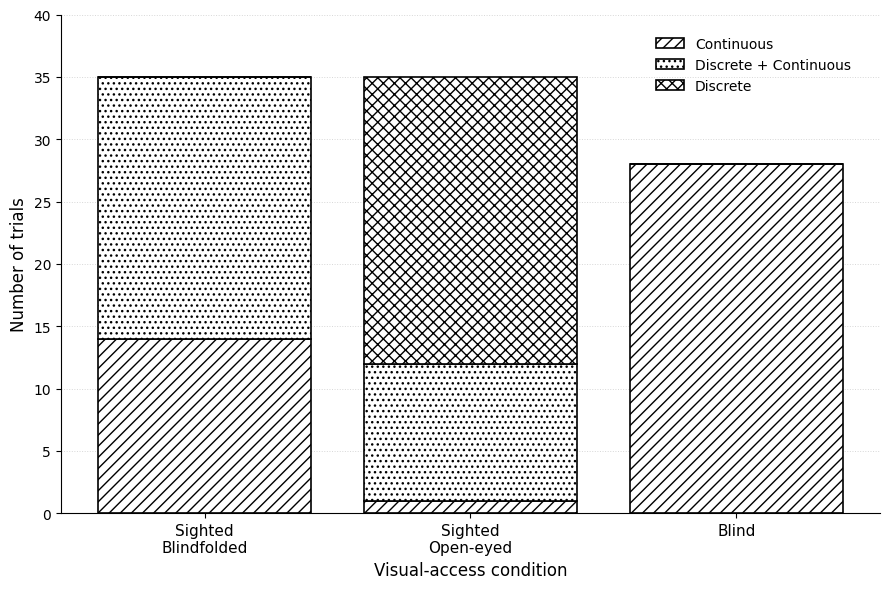}
        \caption{ Surface exploration strategy.}
        \label{fig:contact}
        \Description{The stacked bar chart shows the surface exploration strategies used under three visual-access conditions: Sighted Blindfolded, Sighted Open-eyed, and Blind. The strategies are divided into Continuous, Discrete + Continuous, and Discrete exploration. For the Sighted Blindfolded condition, 14 trials used continuous exploration and 21 trials used discrete + continuous exploration, with no discrete-only trials. For the Sighted Open-eyed condition, 1 trial used continuous exploration, 11 trials used discrete + continuous exploration, and 23 trials used discrete exploration. For the Blind condition, all 28 trials used continuous exploration, with no discrete + continuous or discrete-only trials.}
\end{figure}

As shown in Fig. \ref{fig:contact}, contact behaviour varied systematically with visual access. Sighted participants in the blindfolded condition used discrete and continuous contact strategy, whereas the same patterns with open-eyed experiments showed a marked shift towards discrete contact. This change suggests that visual access allowed the participants to find individual vibrating pixels more directly. In contrast, blind participants consistently used continuous contact across most of the trials which indicates a stronger reliance on movement across the surface to locate and interpret the vibrating regions. The contact behaviour is explored in more detail in Section \ref{sec:rq2}.


\subsection{Related Observations and notes from the field.} 
The index and the middle finger were dominantly used by sighted participants for exploration. Only 1 participant deliberately used the thumb for exploration. The thumb along with the little finger almost always only used for exploration along with the palm. Participants sometimes tried to guess the figure with very limited exploration and the researcher had to prompt them to explore more. One participant took a break after 9 iterations (2 training + 7 blindfolded) and remarked about vibration fatigue setting in with residue sensations in their fingers. Another sighted participant took 5 breaks in the blindfolded study. Their interaction time in each trial was also large. Vertical or columnwise exploration was commonly observed with sighted participants.

\section{Quantitative Analysis}
\label{sec:results}
Each session was video-recorded at 30fps (setup as shown in Fig. \ref{fig:mobileplacement}). To obtain a lightweight proxy for how much a participant's hands moved over time, we reduced each video to a one-dimensional motion-energy signal by frame differencing. Every frame was converted to grayscale, and for each pair of consecutive frames we computed the mean absolute difference in pixel intensity, averaged over all pixels. This yields one value per frame which is near-zero when the scene is static and larger when the hands move, scaling with the amount and speed of movement. Concatenating these values over a trial gives its motion time series, sampled at 30Hz. All subsequent motion analyses (smoothness, rhythmicity, and dwell structure) operate on these signals and are for 98 testing trials.

A response was scored correct only when it exactly matched the presented figure; non-responses were scored incorrect. Because the timing and motion distributions are strongly right-skewed, we report medians and use nonparametric tests (Mann--Whitney $U$; Fisher's exact and exact McNemar for accuracy), with rank-biserial $r$ as the effect size. The eight frame-difference motion features are corrected for multiplicity within their family (Holm). With nine participants the analysis can only identify large effects, if any.

\subsection{RQ1: Rendering fidelity}
\label{sec:rq1}

\begin{figure}[htbp]
  \centering
  \includegraphics[width=0.8\linewidth]{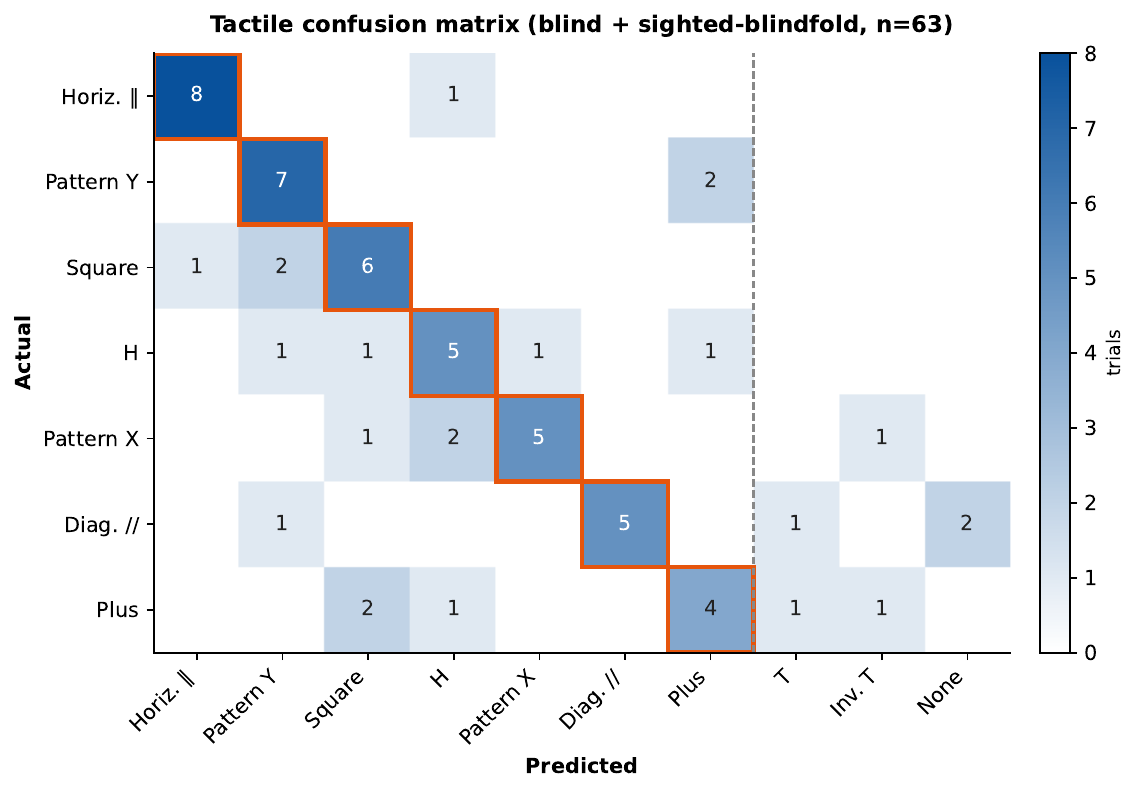}
  \caption{Tactile confusion matrix (blind + blindfolded sighted, $n=63$). Rows
  are presented figures, columns are responses; the three rightmost columns are
  off-target labels and non-responses. Correct cells are outlined. The plus is
  the most confusable target; the diagonal is the only figure to draw
  non-responses.}
  \Description{A heatmap with presented figures on rows and responses on
  columns. Correct (outlined) cells hold the largest counts. The plus row
  spreads into square, H, T and inverted-T; the diagonal row spreads into T and
  the non-response column.}
  \label{fig:confusion}
  \Description{The confusion matrix shows tactile pattern recognition results for the combined blind and sighted-blindfolded participants (n = 63). Rows represent the actual patterns and columns represent the predicted patterns, with the cell values indicating the number of trials. The patterns include Horizontal parallel lines, Pattern Y, Square, Pattern H, Pattern X, Diagonal parallel lines, and Plus, along with Pattern T, Inverted T, and None as predicted responses. The diagonal cells show 8 correct responses for Horizontal parallel lines, 7 for Pattern Y, 6 for Square, 5 for Pattern H, 5 for Pattern X, 5 for Diagonal parallel lines, and 4 for Plus. The remaining values represent misclassifications across the different patterns, including responses classified as T, Inverted T, or None.}
\end{figure}

Table \ref{tab:pattern_accuracy} shows the overall accuracy. Recognition on purely tactile trials (blind and blindfolded sighted pooled,
$n=9$ per figure) varied widely across figures (Table~\ref{tab:perfigure}).
Horizontal parallel lines were almost always identified (89\%), whereas the plus
was hardest (44\%). The confusion structure (Figure~\ref{fig:confusion}) is
informative. The \emph{plus} was mislabelled as square, T, inverted-T, or
H, each a sub-shape or relative of a cross consistent with incomplete radial
exploration that recovers only part of the figure. The angular step-figures
(Pattern~X, Pattern~Y) were confused with their rectilinear neighbours (H,
square) that share edges and corners.

\emph{Diagonal parallel lines} were a distinct failure of the rendering itself.
The figure drew four of the study's five non-responses; two of these occurred
\emph{with the eyes open}, where sighted participants could not identify it even while looking at it. Diagonal was also the slowest figure to
resolve (median 145\,s) and the only figure whose accuracy fell from the
blindfolded to the open-eyed stage. This means that the diagonal's encoding in the surface and/or in the reference sheet is the source of difficulty.

\begin{table}[htbp]
  \caption{Tactile identification accuracy per figure (blind + blindfolded
  sighted, $n=9$ each), with median resolution time.}
  \label{tab:perfigure}
  \small
  \begin{tabular}{lcc}
    \toprule
    Figure & Accuracy & Median time (s) \\
    \midrule
    Horizontal parallel lines & 89\% (8/9) & 58 \\
    Pattern Y (see Fig. \ref{full})        & 78\% (7/9) & 90 \\
    Square                    & 67\% (6/9) & 83 \\
    H                         & 56\% (5/9) & 81 \\
    Pattern X (see Fig. \ref{full})          & 56\% (5/9) & 61 \\
    Diagonal parallel lines   & 56\% (5/9) & 145 \\
    Plus                      & 44\% (4/9) & 68 \\
    \bottomrule
  \end{tabular}
\end{table}

\subsection{RQ2: Exploration strategy}
\label{sec:rq2}

\begin{figure}[htbp]
  \centering
  \includegraphics[width=\linewidth]{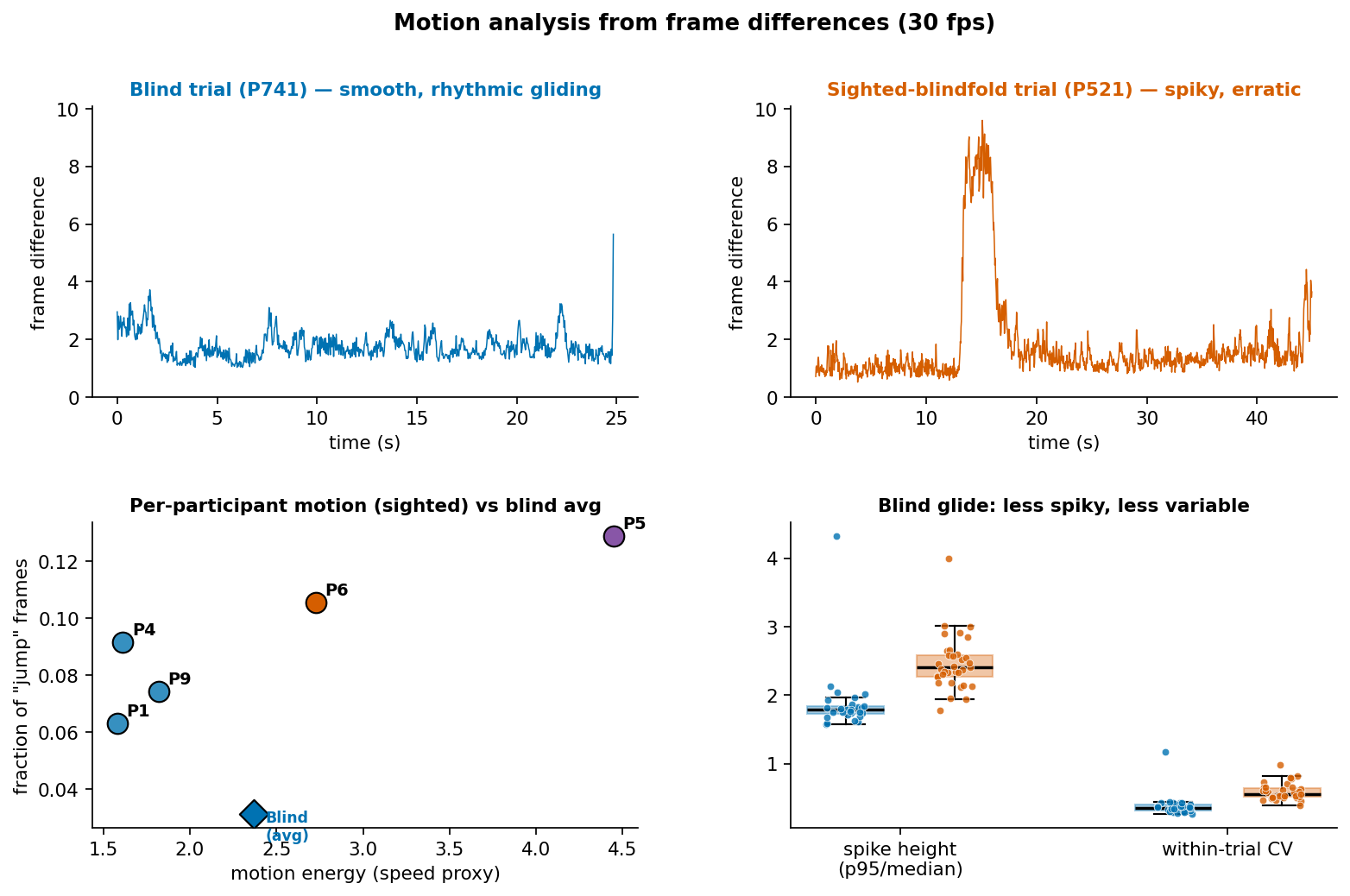}
  \caption{Motion from frame differences (30\,fps). Top: example traces---a blind
  trial (smooth, rhythmic gliding) and a sighted-blindfolded trial (spiky,
  erratic). Bottom left: per-participant motion energy vs.\ jump-frame fraction
  for sighted participants, against the blind average. Bottom right: blind trials
  are less spiky and less variable within a trial.}
  \Description{Top row: two time-series; the blind trace is a steady
  low-amplitude oscillation, the sighted trace is quiet then bursts to large
  values. Bottom left scatter: sighted participants spread to higher motion and
  jumpiness than the blind average. Bottom right box plots: blind lower on spike
  height and within-trial variability.}
  \label{fig:motion}
  \Description{The figure shows motion analysis based on frame differences at 30 fps. The top-left plot presents an example blind trial (P741), showing relatively smooth and rhythmic motion over approximately 25 s. The top-right plot presents a sighted-blindfolded trial (P521), showing a pronounced spike in frame difference around 14–16 s followed by more variable motion. The bottom-left scatter plot compares motion energy with the fraction of jump frames for sighted participants (P1, P4, P5, P6, and P9) and the blind average. The bottom-right plot compares the two groups using spike height (p95/median) and within-trial coefficient of variation (CV), showing lower spike height and variability for the blind trials.}
\end{figure}

\paragraph{Two reading strategies.}
Frame-difference signals separated two exploration styles---a continuous
\emph{glide} and a \emph{halting} search---that mapped onto the two participant groups (Figure~\ref{fig:motion}). Relative to blindfolded sighted participants, blind participants' motion was smoother (within-trial coefficient of variation 0.36 vs.\ 0.55), less spiky (95th-percentile/median 1.79 vs.\ 2.41), and less prone to large jumps (0.03 vs.\ 0.09 of frames), and its movement spectrum was more concentrated (higher spectral peakedness, lower flatness; all five measures Holm-corrected $p<.002$; largest $r=.91$).

\begin{figure}
    \centering
    \includegraphics[width=0.5\linewidth]{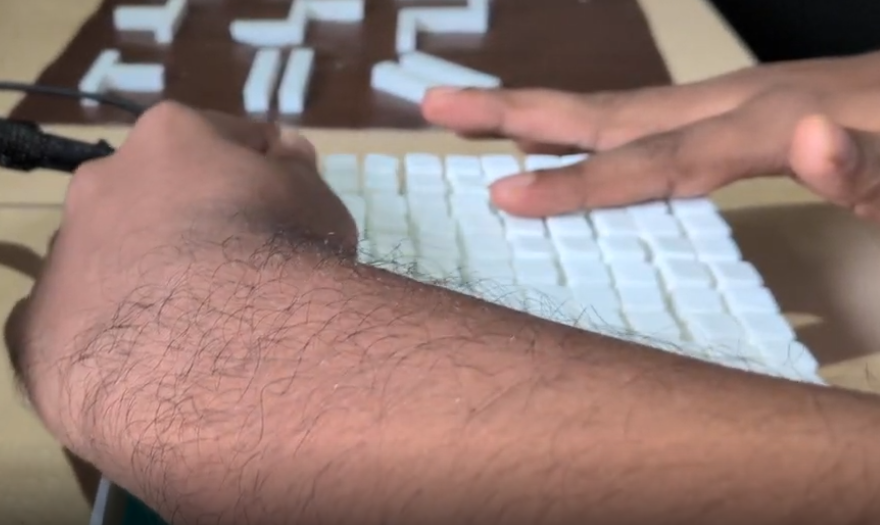}
    \caption{The screenshot corresponding to the spike in the sighted-blindfold trial P521 in Figure \ref{fig:motion}. The spike is a result of the user switching their left-hand from a vertical or column-scanning strategy (as shown from their right hand) to a horizontal or a row-scanning strategy.}
    \label{fig:erratic}
    \Description{The image shows a participant exploring the vibrotactile surface using both hands during a sighted-blindfolded trial. The participant’s left hand is positioned over the surface while the right hand moves across the pixels, illustrating a change in exploration direction from vertical or column-wise scanning to horizontal or row-wise scanning. The tactile patterns are visible in the background of the setup.}
\end{figure}

\begin{figure}[htbp]
  \centering
  \includegraphics[width=\linewidth]{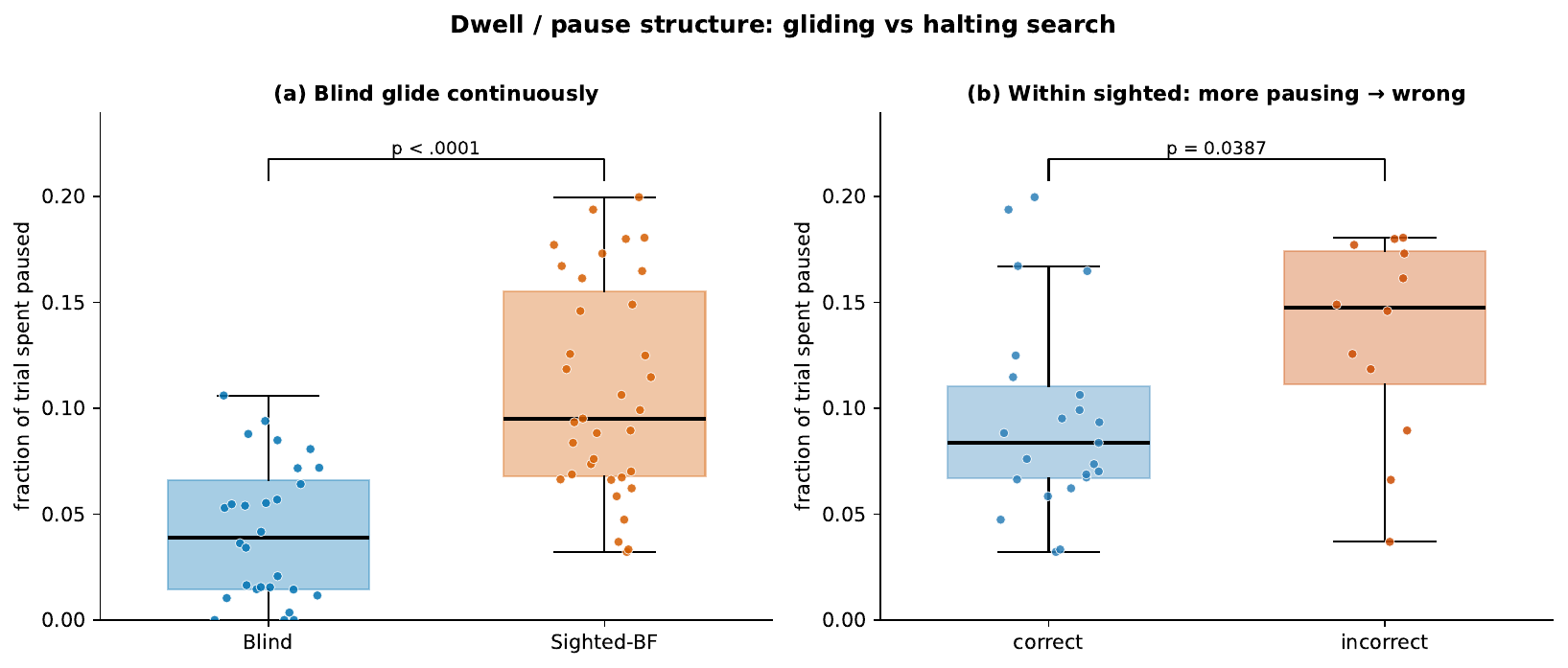}
  \caption{Dwell/pause structure. (a) Blind participants spend far less of a trial
  paused than blindfolded sighted participants. (b) Within the sighted group,
  incorrect trials contain more pausing than correct ones. Boxes show median and
  IQR; points are trials.}
  \Description{Two box-and-strip plots. Panel (a): fraction of the trial paused is
  much lower for Blind than Sighted-Blindfold, significant below .0001. Panel (b):
  within sighted, incorrect trials have a higher paused fraction than correct
  trials, p about .04.}
  \label{fig:dwell}
  \Description{The figure shows the dwell and pause structure of surface exploration across two comparisons. In (a), the fraction of trial time spent paused is compared between blind and sighted-blindfolded participants; blind participants show a lower median fraction of paused time, while sighted-blindfolded participants show a higher median. The difference is statistically significant (p < 0.0001). In (b), trials from the sighted-blindfolded group are divided into correct and incorrect responses. Incorrect trials show a higher median fraction of paused time than correct trials, although the difference is not statistically significant (p = 0.0387). The box plots show the median and interquartile range, with individual points representing trials.}
\end{figure}

\paragraph{Continuous glide vs.\ stop-and-start.}
The clearest expression of this difference is in \emph{pausing}
(Figure~\ref{fig:dwell}a). Defining a pause as a run of low-motion frames (below
$0.7\times$ the trial's median, lasting $\ge0.2$\,s), blind participants paused
about half as often (0.13 vs.\ 0.24 pauses/s) and spent roughly $2.4\times$ less
of the trial stationary (4\% vs.\ 10\%; $U=132$, $p<.0001$, $r=.73$; bootstrap
95\% CI on the median difference $[-0.10,-0.03]$). The contrast holds across
amplitude thresholds (0.5--0.8) and minimum pause durations (0.10--0.50\,s).
Blind participants read largely {through} continuous motion: a pattern consistent with the continuous-fingertip motion used in tactile (e.g.\ Braille) reading, which the blind participants may be transferring. On the other hand, sighted participants fell back on a discontinuous probe-and-hold search--- the pixelated design of the surface might have resulted in a transfer of interaction patterns with a computer keyboard (see for example, Figure\ref{full}). 

\paragraph{Relation to success and time.}
More pausing was associated with \emph{worse} recognition, but only among the sighted (Figure~\ref{fig:dwell}b): sighted-blindfolded trials that were incorrect contained more pausing than correct ones (fraction paused 0.15 vs.\ 0.08; $U=78$, $p=.039$, $r=.43$; CI $[-0.10,-0.02]$), whereas within the blind group pausing was unrelated to accuracy ($p=1.0$). Reading \emph{time}, by contrast, did not correlate to success at all: correct and incorrect tactile trials took statistically indistinguishable time (median 76\,s vs.\ 81\,s; $p=.81$). Persistence did not characterize successful reading.

\paragraph{Use of the reference sheet.}
Participants also differed in how they combined the surface with the reference
sheet (Table~\ref{tab:allocation}). Blind participants spent 35\% of their time
exploring the surface and the reference \emph{simultaneously}---one hand on
each---whereas blindfolded sighted participants did so for only 7\%, consulting
the reference in a separate pass. 

\paragraph{An illustrative failure.}
Participant P6 scored 0/7 blindfolded yet recovered to 5/7 open-eyed on
the same figures---a tactile-decoding failure despite an otherwise normal strategy and full engagement.

\begin{table}[htbp]
  \caption{Share of time on the surface alone, the reference sheet alone, and
  both simultaneously.}
  \label{tab:allocation}
  \small
  \begin{tabular}{lccc}
    \toprule
    Condition & Surface & Sheet & Both (simultaneous) \\
    \midrule
    Blind             & 57.8\% & 7.1\% & \textbf{35.1\%} \\
    Sighted-Blindfold & 86.1\% & 6.7\% & 7.2\% \\
    Sighted-Open-Eye  & 98.5\% & 1.5\% & 0.0\% \\
    \bottomrule
  \end{tabular}
\end{table}

\subsection{RQ3: Effect of visual access}
\label{sec:rq3}

For sighted participants, adding vision sharply reduced reading time: the median
total time per trial fell from 135\,s blindfolded to 65\,s open-eyed
(Figure~\ref{fig:timing}a; $U=956$, $p<.0001$, $r=.56$). Its effect on accuracy
was smaller and not statistically detectable at this sample size: accuracy rose
from 66\% (23/35) to 80\% (28/35), but the paired comparison was not significant
(exact McNemar on 7 gains vs.\ 2 losses, $p=.18$). Vision's benefit here is
primarily one of speed.

Not all confusions were resolved by sight. Most figures improved or held from the
blindfolded to the open-eyed stage (e.g.\ the plus, $3/5\rightarrow5/5$), but
\emph{diagonal parallel lines got worse} ($2/5\rightarrow1/5$) and the square did
not improve (3/5 in both). 

\subsection{RQ4: Blind vs.\ blindfolded sighted touch readers}
\label{sec:rq4}

\begin{figure}[t]
  \centering
  \includegraphics[width=\linewidth]{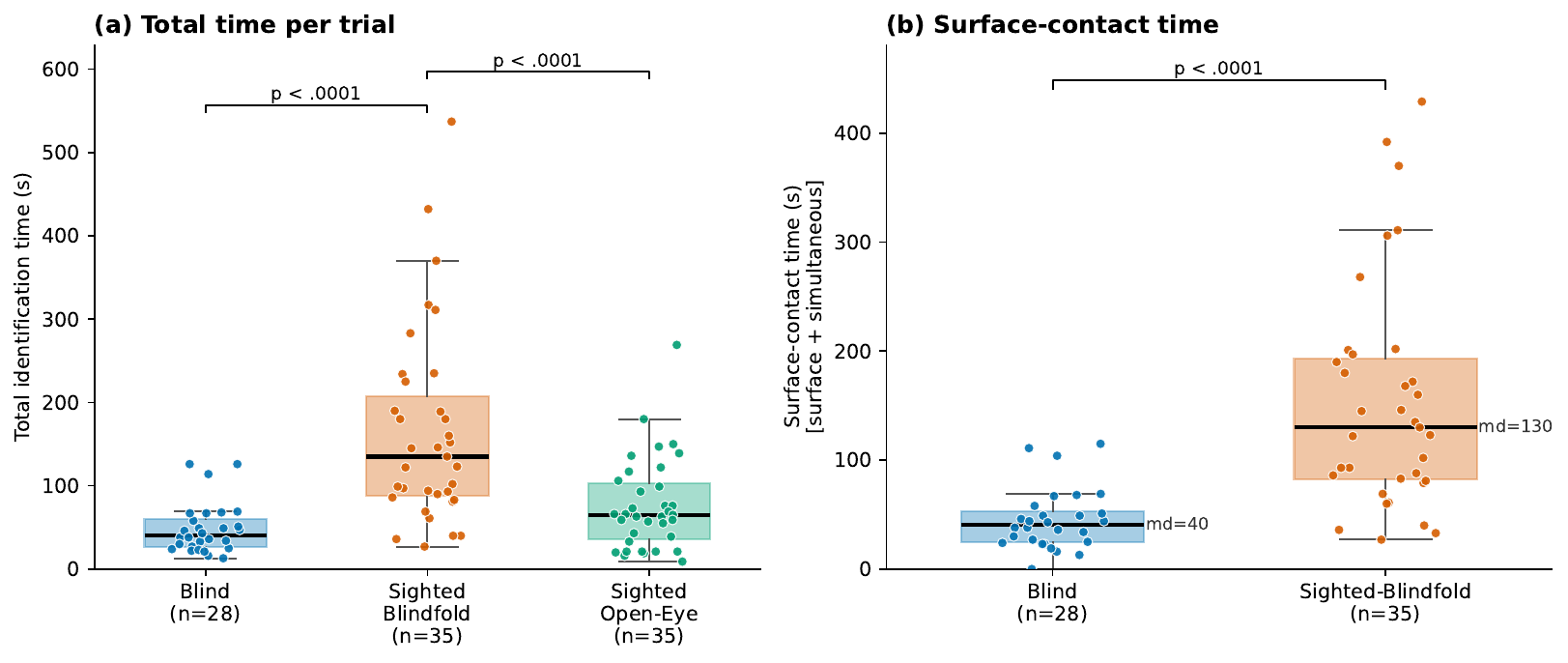}
  \caption{Identification time by condition. (a) Total time per trial: blind
  participants are far faster than blindfolded sighted participants and faster
  even than sighted participants with vision. (b) Surface-contact time
  (surface-alone plus simultaneous surface-and-sheet), blind vs.\ blindfolded
  sighted. Boxes show median and IQR; points are individual trials.}
  \Description{The figure compares identification time across three visual-access conditions. In (a), total identification time per trial is shown for Blind (n = 28), Sighted Blindfolded (n = 35), and Sighted Open-Eye (n = 35). Blind participants have the lowest median identification time, while sighted-blindfolded participants show the highest median, with significant differences between the conditions (p < 0.0001). In (b), surface-contact time, including surface-only and simultaneous surface-and-sheet contact, is compared between blind and sighted-blindfolded participants. The median surface-contact times are approximately 40 s for blind participants and 130 s for sighted-blindfolded participants, with a significant difference (p < 0.0001). Boxes represent the median and interquartile range, and individual points represent trials.}
  \label{fig:timing}
\end{figure}

Reading purely by touch, blind and blindfolded sighted participants achieved
statistically indistinguishable accuracy (61\%, 17/28, vs.\ 66\%, 23/35;
Fisher's exact $p=.79$). They differed sharply, however, in \emph{speed}. Blind
participants identified a figure in a median of 40\,s against 135\,s for
blindfolded sighted participants---a large, highly significant difference
(Figure~\ref{fig:timing}a; $U=106$, $p<.0001$, $r=.78$). Counting all time in contact with the surface (including simultaneous surface-and-sheet exploration), blind participants spent a median of 40\,s versus 130\,s (Figure~\ref{fig:timing}b; $p<.0001$, $r=.78$).

This efficiency reflects the difference established in Section \ref{sec:rq2}: the two groups embody the two reading strategies. Blind participants glide continuously and cross-reference the sheet concurrently, while blindfolded sighted participants search in halting, pause-heavy passes---at equal accuracy but roughly three times the cost. Confusion structure was broadly similar across the two
groups (Figure~\ref{fig:confusion} pools them), with the plus and diagonal
difficult for both.

\subsection{Summary}
The surface renders simple axis-aligned line figures reliably but not the plus or
the diagonal, the latter a representation problem that vision does not fix (RQ1,
RQ3). Reading proceeds through one of two strategies---continuous rhythmic gliding
or halting probe-and-hold---and, among sighted readers, the halting mode co-occurs
with errors while time-on-task does not predict success (RQ2). Vision mainly buys
speed, not accuracy (RQ3). Reading purely by touch, blind participants match
blindfolded sighted participants on accuracy but are about three times faster, by
virtue of the gliding, concurrently-referenced strategy (RQ4).

 
\section{Discussion}
\label{sec:discussion}
 
Read purely by touch, a low-cost vibrotactile surface let participants identify simple figures well above chance, and blind and blindfolded-sighted readers reached comparable accuracy. What separated them was not
whether they could read the surface but \emph{how}. Two themes emerge---one about the reader, one about the display---and together they motivate design implications for affordable refreshable tactile displays.

\subsection{Design implications for low-cost refreshable tactile displays}
\label{sec:disc-implications}
 
Taken together, our results suggest concrete guidance for affordable refreshable
tactile displays and for how figures should be represented on them. We state these
as implications supported by our findings rather than as validated design rules.
 
\paragraph{Invest in representation before resolution.}
Simple axis-aligned figures (horizontal lines) were read reliably, whereas the plus was confused with its own sub-shapes and the diagonal was misidentified even with the eyes open. Limited budget (for affordability) is better spent on encodings that degrade gracefully at low resolution than on additional actuators. Favour figures whose defining features are axis-aligned and separable; treat oblique lines, fine crossings, and features that alias against the grid as costly, and render them with deliberate, disambiguating encodings or multi-modal support.
 
\paragraph{Support continuous, uninterrupted gliding.}
Since effective reading proceeds \emph{through} continuous motion, the physical surface must let a fingertip glide
without catching. Flush mounting, consistent actuator feel, and mechanical
reliability are not polish but core to legibility. In our study, actuators that
protruded, dislodged, or made the casing shake both interrupted gliding and, at
times, leaked non-tactile cues about which elements were active.
 
\paragraph{Guarantee per-element perceptibility and reliability.}
It is possible that during exploration, some of the motors became temporarily inactive which might have led to long exploration times or incorrect recognition. The present design requires mechanical stability for electrical contact. Reliable, well-above-threshold actuation matters more than the number of elements.

 
\paragraph{Design for sustained use.}
Reading appeared to degrade after several minutes of continuous vibration due to persistent sensation in users fingers. Affordable displays built on vibration motors should anticipate tactile adaptation through duty-cycling, adaptive intensity, or paced interaction so that patterns remain legible across a session.
 
\paragraph{Onboard novices toward gliding, and keep the reference in the loop.}
Because the efficient strategy is learned rather than innate to the device, brief
guidance encouraging continuous scanning (rather than static probing) and
systematic anchoring may help novice and sighted users approach the fluency blind
readers showed. Blind participants also interleaved the surface and the reference
legend continuously; keeping an easily accessible reference within the reading
loop, rather than as a separate lookup, fits how skilled users already work.
 
\subsection{Limitations}
\label{sec:limitations}
 
These implications must be read against the study's limitations, which also bound
what our findings can claim. This is an exploratory, mixed-methods study; its
contribution is to characterize how figures on this surface are read and to surface
design implications, not to establish population-level effects.
 
\paragraph{Sample and population.}
Nine participants (four blind, five sighted) each contributed seven tactile trials. The generality of the results is not established. 
On the other hand, because tactile-reading expertise is central to our findings, the characteristics of our blind participants matter and the results may not transfer to blind users with different Braille literacy or prior experience with tactile displays.
 
\paragraph{Global motion measure.}
Our motion analysis derives from frame differencing, which yields a single
motion-energy value per frame, aggregating both hands, the reference sheet, and any background movement. It captures how much and how rhythmically movement occurred, but not where on the surface, which hand, or in what direction. Spatial analysis of coverage, per-hand trajectories, scanning direction would require hand or pose tracking on the raw video. Camera positioning further constrained the analysis: in some trials the exploring hand partially occluded the surface, obscuring fine movements and forcing the coder to infer them.
 
\paragraph{Coding and apparatus.}
Contact-mode and finger-count codes record the presence of contact, not sustained
exploration, and should be read as breadth of contact rather than effort; contact
mode was double-coded with moderate agreement (Cohen's $\kappa = 0.60$), while the
remaining field observations reflect a single coder's interpretation. The display design itself introduced confounds. Mechanical movement due to interaction might have resulted in break in electrical contact of the motors. Possibly weak or inactive motors could mean figures were sometimes
reconstructed from incomplete stimulation. In addition, shaking casing offered an incidental visual cue as to which elements were active. Finally, sighted participants always completed the blindfolded stage before the open-eyed stage, so open-eyed performance may reflect familiarity with the figures in addition to visual access; the two cannot be separated in this design.
 
\subsection{Future Work}
\label{sec:futurework}
 
The limitations above, together with the design directions they suggest, point to
several next steps spanning the display, the study design, and the analysis.
 
\paragraph{Representation and sustained-use redesign.}
The clearest device-level result is that some figures fail at the level of
representation rather than sensing. Future work should treat figure representation
as a design variable---systematically varying line rendering, motor density, and
how oblique features are approximated on a discrete grid---and measure the
resulting legibility. The apparent decline after prolonged exposure similarly
motivates studying the time-course of tactile adaptation on the display, and
mitigations such as duty-cycling or adaptive intensity.
 
\paragraph{Powered, counterbalanced replication.}
Because our sample supports only large effects, the accuracy comparisons remain
open. The effect sizes we report motivate a larger confirmatory study, analysed
with mixed-effects models that account for the repeated-measures structure, that
counterbalances the blindfolded and open-eyed stages to separate vision from
familiarity, and that stratifies blind participants by tactile-reading background
so expertise can be tested as a moderator.
 
\paragraph{Testing the tactile-expertise account.}
We interpreted the continuous ``gliding'' strategy and the threefold speed
advantage as transfer from tactile reading such as Braille. This can be tested
directly by comparing Braille-literate and non-Braille-reading participants: if the
gliding signature tracks Braille literacy, the transfer hypothesis is supported; if
sighted readers converge toward gliding with training, the strategy is learnable
rather than expertise-bound---either outcome informing how new users are onboarded.
 
\paragraph{Richer instrumentation and content.}
Adding more cameras for motion campus with hand and pose tracking would recover the spatial features such as coverage, per-hand trajectories, scanning direction, and surface-to-reference switches and would let the concurrent surface-and-reference strategy be measured directly. Finer coding that distinguishes sustained exploration from incidental contact, and a reading dwell from a deliberative pause, would sharpen the strategy-accuracy link. Once static figures render reliably, the display should be evaluated on larger symbol sets, dynamic patterns, and functional tasks such as reading charts, maps, or diagrams.

\paragraph{Repairability} A user-study with Blind participants on the unaided repairability of this design.


\section*{Generative AI Usage}

Generative AI tools were used during the preparation of this
manuscript for limited assistance with text refinement and image
generation (Fig.~\ref{fig:mobileplacement}). All AI-assisted
content was reviewed, verified, and edited by the authors before
inclusion. No confidential, proprietary, personally identifiable,
or sensitive research data were provided to any generative AI
system. The authors remain fully responsible for the accuracy,
integrity, and final content of the manuscript.

\bibliographystyle{ACM-Reference-Format}
\bibliography{reference}


\section{Appendix}



\subsection{Measuring Vibration Spread} 
We used an MPU-6050 accelerometer mounted on a breakout board to measure the vibration. The accelerometer is placed directly on the appropriate pixel for measurement. Data is collected at a sampling rate $f_s = 1\,\text{kHz}$ for $T = 2\,\text{s}$, producing $N = f_s \, T$ samples for each sub-pixel. The sensor provides three-axis accelerations ($a_{x},a_{y},a_{z}$). For the $n$-th sample, let $a_{\text{mag}}(n) = \sqrt{a_x(n)^2 + a_y(n)^2 + a_z(n)^2}$. To cover for fabrication and assembly imperfections, for each sub-pixel, we first record the baseline acceleration with the appropriate motors OFF ($a_{\text{mag}}^{OFF}(n)$) followed immediately by the measurements with the appropriate motors ON ($a_{\text{mag}}^{ON}(n)$). The offset-corrected magnitude (per sample) is $\Delta a_{\text{mag}}(n)$ and we characterize the vibration in a sub-pixel by the RMS value $a_{rms}$.
\begin{equation}
    \Delta a_{\text{mag}}(n) = a_{\text{mag}}^{ON}(n) - a_{\text{mag}}^{OFF}(n); \quad \quad a_{rms}= \sqrt{\frac{1}{N}\sum_{n=1}^{N} \Delta a_{\text{mag}}(n)^2} \nonumber
\end{equation}

\end{document}